\documentclass[a4paper,fleqn,usenatbib]{mnras}

\usepackage{mathptmx}
\usepackage[T1]{fontenc}
\usepackage{ae,aecompl}

\usepackage{graphicx}	% Including figure files
\graphicspath{{./plots/}}
\usepackage{amsmath}	% Advanced maths commands
\usepackage{amssymb}	% Extra maths symbols
\usepackage{epstopdf}

\usepackage{color}

\newcommand{\solarmass}{\text{M}_{\odot}}
\newcommand{\Myr}{\text{Myr}}
\newcommand{\Gyr}{\text{Gyr}}

\title[Ram-Pressure Stripping on Dwarf Galaxies]
{The Effect of Ram-Pressure Stripping on Dwarf Galaxies}

\author[Steyrleithner et al.]{
P. Steyrleithner $^{1}$,\thanks{E-mail: patrick.steyrleithner@univie.ac.at}
G. Hensler $^{1}$,
A. Boselli $^{2}$
\\
$^{1}$Department for Astrophysics, University of Vienna, T\"urkenschanzstrasse 17, A-1180 Vienna, Austria\\
$^{2}$Aix Marseille Universit\ae, CNRS, LAM (Laboratoire diAstrophysique de Marseille), UMR 7326, F-13388 Marseille, France 
}

\date{Accepted XXX. Received YYY; in original form ZZZ}

\pubyear{2018}

\begin{document}
\label{firstpage}
\pagerange{\pageref{firstpage}--\pageref{lastpage}}
\maketitle

\begin{abstract}
Ram-pressure stripping (RPS) is a well observed phenomenon of massive spiral galaxies passing through the hot intra-cluster medium (ICM) of galaxy clusters. For dwarf galaxies (DGs) within a cluster, the transformation from gaseous to gas-poor systems by RPS is not easily observed and must happen in the outskirts of clusters. In a few objects in close by galaxy clusters and the field, RPS has been observed. Since cluster early-type DGs also show a large variety of internal structures (unexpected central gas reservoirs, blue stellar cores, composite radial stellar profiles), we aim in this study to investigate how ram pressure (RP) affects the interstellar gas content and therefore the star-formation (SF) activity. Using a series of numerical simulations, we quantify the dependence of the stripped-off gas on the velocity of the infalling DGs and on the ambient ICM density. We demonstrated that SF can be either suppressed or triggered by RP depending on the ICM density and the DGs mass. Under some conditions, RP can compress the gas, so that it is unexpectedly retained in the central DG region and forms stars. When gas clouds are still bound against stripping but lifted from a thin disk and fall back, their new stars form an ellipsoidal (young) stellar population already with a larger velocity dispersion without the necessity of harassment. Most spectacularly, star clusters can form downstream in stripped-off massive gas clouds in the case of strong RP. We compare our results to observations.
\end{abstract}


%%%%%%%%%%%%%%%%%%%%%%%%%%%%%%%%%%%%%%%%%%%%%%%%%%

%%%%%%%%%%%%%%%%% BODY OF PAPER %%%%%%%%%%%%%%%%%%

\section{Introduction}

Dwarf galaxies (DGs) are the most numerous type of galaxies in the universe and have low surface brightnesses, low gravitational potential wells, and relatively low star-formation rates (SFRs). Due to their low gravitational potential wells, they react sensitively to external and internal processes such as feedback by supernovae (SNe), and are therefore ideal objects to study galaxy evolution. 

During their evolution, galaxies suffer transformations from gas-rich to gas-poor; this is particularly true for DGs. 
The chance to be morphologically transformed is larger in high density regions. The morphology-density relation \citep{Dressler80} describes the morphology of a galaxy as a function of the projected galaxy number density within a galaxy cluster.
The higher the projected density is, the lower the fraction of gas-rich galaxies, and the higher is the fraction of early-type galaxies. In galaxy clusters, DGs as gas-poor elliptical galaxies (dEs) outnumber all other galaxy types \citep{Binggeli88}. 

Several studies on DGs in the Virgo cluster by \cite{LGB06,LGW06,LJH09} found that dEs are much more complex and can be divided into morphological subclasses such as dEs with disk features, dE(di), with a blue centre, dE(bc), (non)nucleated, dE(nN)/dE(N), and normal dEs. 
Further detailed studies by the SMAKCED survey in \cite{Janz12} discovered various compositions of the stellar brightness profiles.
Some of these dE(bc)s also show a significant rotation, which supports the theory that they are the descendants of gas-rich star-forming dwarf irregular galaxies (dIrrs), which are known to be mostly rotationally supported systems. \cite{Tol15} also found that slow-rotating dEs are more roundish and move on more cluster-centric orbits, whereas fast rotators are flat \citep{LJH09} and exist at larger Virgo-centric distances. In addition, fainter DGs are systematically thicker \citep{Sanchez19}.
Moreover, the radial cluster distribution of flat dEs obtrudes their recent cluster infall \citep{LJH09}. 
\cite{LGW06} performed a spectral analysis of these dE(bc)s and found that these DGs show Balmer line emissions and some H$\alpha$ detections, which are in this region a possible sign of active SF.   
  
It is important to understand how rotationally flattened late-type star-forming dIrrs are transformed by the cluster environment into roundish early-type DGs that can partly retain their gas. As reviewed by \cite{BG14}, two main mechanisms can be invoked to extract the gas from gas-rich galaxies. The first is harassment, which is an effect of a high-speed encounter of a dIrr with a massive galaxy, which increases not only the stellar velocity dispersion, making the axial ratio more spherical \citep{Smith15}, but also disrupts the gas disk \citep{MMM05,Bia15}. This mechanism also predicts that the gas gets funnelled to the centre where it may form stars. The second mechanism is ram-pressure stripping (RPS) by the ICM, which removes the interstellar medium (ISM) from the DG \citep{BBC08b,BBC08a}. The RP acts on the gas of a galaxy from its leading side when falling into the cluster. This can trigger multiple effects, including the compression of cold gas in the centre and the resulting formation of new stars, as has been seen in numerical simulations, e.g. \cite{RBO14}. 

The dominance of blue-cored dEs in the closest galaxy clusters \citep{Urich17,Hamraz19} indicated that recent SF in the centres of DGs has taken place. The hot gas and less bound circum-galactic gas is stripped off first and when this gas reservoir is exhausted, SF will be quenched due to the cessation of further gas infall. 
Further stripping of a dE(bc)s will unavoidably evolve them into normal dEs with red colours. When RPS has transformed dIrrs to dEs, these must keep the rotationally-supported flat stellar disk and be distinguished as those \citep{Zee04}. In order to thicken the surviving thin stellar disk after RPS gas loss to an ellipsoidal shape as additional heating of the stellar velocity dispersion harassment is further on necessary \citep{Smith15}.

There are two major phases of RPS. The instantaneous stripping phase happens when a galaxy moves with a relative velocity through the ICM and RPS will affect at first the low density gas since it has a low gravitational binding. The outer gas disk is stripped off from outside-in, until a certain minimal stripping radius is reached. The second phase is the continuous stripping phase. Galaxies that survive the instantaneous stripping phase will lose their gas subsequently due to Kelvin-Helmholtz instabilities occurring at the interface between the gas of the galaxy and that of the ICM \citep{RH05}. 

\cite*{GG72} showed that a spiral galaxy loses its gas when and where the RP exceeds the restoring gravitational force that binds the gas to the galaxy. This condition is given by $P_{ram} > 2\pi\, G \ \Sigma_s\,\Sigma_g$, where $\Sigma_s$ and $\Sigma_g$ are the star and gas surface density, respectively. DGs have smaller escape velocities and they are therefore more susceptible to RPS than more massive galaxies. 
\cite{RH05} demonstrated in their simulations the effect of RP on mass loss of massive galaxies moving through the ICM and found a third stripping phase intermediate between the instantaneous and the continuos stripping phases. This is a dynamical phase where a part of the pushed out material is stripped completely and the rest falls back onto the leeward side of the galaxy in their wind shadow.
\cite*{MB00}, on the other hand, have compared the RP with the thermal pressure at the centre of the gravitational potential wells of the dark matter (DM) halos of DGs and showed the importance of DM halos for the amount of the final gas mass loss by RPS. 

\cite{MBD03} investigated the inclination angle at which the DG moves through the ICM and the effect of RP on the gas loss of DGs. 
Before getting stripped off, the gas is compressed by RP which results in an increase of SF, and both the strengths and spatial distributions of this additional SF are currently unknown. 

The effect of RPS on the (hot) gas is well observed in massive gas-rich galaxies. For example, NGC 4569 in the Virgo cluster serves as a prototypical example in which not only stripped-off neutral gas is detected \citep{Vol04}, but recently also faint filamentary structures of ionised gas \citep{BCF16}. This and other RPS-affected spirals are at first detected in the Virgo cluster where they pass through the hot ICM.  

For DGs, this RPS signature is challenging to observe due to their low surface brightnesses and because their shallow potential wells allow for easier gas loss. 
According to the RP formula by \cite{GG72}
\begin{align}\label{eg:Pram}
	P_{ram} = \rho \, v^2 \, 
\end{align}
where $\rho$ is the ICM density and $v$ the velocity of the infalling galaxy relative to the ICM at rest. DGs with v = 1000 km s$^{-1}$ and a total mass below $10^9\, \solarmass$ should lose all their gas by means of RPS when passing through the Virgo cluster ICM \citep[priv. commun. with T. Lisker]{BBC08a}. In contrast, remaining gas in DG centres leading to star-forming and blue cores challenge this understanding \citep{LGB06}. 
If one wants to catch a DG during RPS, it depends on the right location. 
Observations show that DGs in galaxy clusters (e.g. Virgo) can be transformed morphologically from late-type to early-type galaxies by various external effects \citep{LGB06, LGW06, LJH09}. By an analysis of the morphological transformation of DGs in the Fornax cluster, \cite{DVB10} showed that the observed morphology-density relation of infalling dwarfs is mainly due to the effect of RPS. 
In the last decade, several DGs with ram-pressure stripped tails were observed within clusters \citep{HSN10, FGS11, YGF13, JKR13, JCC14, KGJ14, JKK15}. One of them is the low-surface brightness DG IC 3418, which is caught during the transformation via RPS. IC 3418 shows a $17\,\text{kpc}$ bright UV tail with bright knots, indicating recent or ongoing SF, while the SF in the main body has ceased some time ago. \cite{KGJ14} identified 10 distinct UV sources (diffuse regions, knots, linear streams and head-tail) and eight of them also exhibit HII regions in the kernels of head-tail structures. 

Morphological transformations of DGs by RPS does not only happen in rich galaxy clusters but also in galaxy groups (GGs) where relative velocities and inter-galactic gas densities are lower. Since galaxy clusters accumulate mass also by the infall of GGs \citep{Lis18}, DGs can be already pre-processed when they enter the cluster environment. 

This paper is structured as followed: section \ref{sec:simulation} provides the simulations ingredients, section \ref{sec:numsetup} provides the numerical setup, section \ref{sec:models} describes the DGs model, in section \ref{sec:results} the results are presented which are discussed in section \ref{sec:discussion}, and finally the 
conclusions in section \ref{sec:conclusion}.

%%%%%%%%%%%%%%%%%%%%
\section{Ingredients of the Simulations}\label{sec:simulation}

\subsection{The Code}

In this work we use the simulation code cdFLASH, which is based on the FLASH code version 3.3 \citep{Fryxell00}. The FLASH code provides the user with several operator splitting methods for treating the hydrodynamics, such as the third-order piecewise-parabolic method (PPM) which uses the second-order Strang time splitting. Besides the direct split methods there are also direct unsplit methods available in FLASH v3.3. We use the unsplit Monotonic Upstream-centred Scheme for Conservation Laws (MUSCL) Hancock solver, which has a second-order accuracy in space and time, with the HLLC Riemann solver and the van Leer slope limiter. To advance the simulation on the minimum time step, the Courant-Friedrich-Lewy (CFL) condition \citep{CFL67} is used. For the simulations presented here, the CFL constant is set to 0.1. This value although much lower than for pure hydrodynamical simulations necessary makes the numerical modelling more time expensive, but can guarantee to prevent unphysical negative densities which can emerge for high density and velocity contrasts at interfaces and to account for cooling and SF even when timescales are temporarily shorter than the dynamical one. \\
The highest spatial resolution is achieved according to
\begin{align}
	\Delta x = \frac{D}{n_{cells,block} \, 2^{l_{ref}-1}} \, ,
\end{align}
with $D$ as the size of the simulation box, $n_{cells,block}$ the number of cells per block and $l_{ref}$ the maximum level of refinement. 
For all simulations we use $D$ of $\pm 12.8\,\text{kpc}$, $n_{cells,block}$ is set to 16 and $l_{ref}$ to 6. This leads to a maximal resolution of $50\,\text{pc}$.

%%%%%%%%%%%%%%%%%%%%
\subsection{Initial Conditions}\label{subsec:IC}

\subsubsection{The mass distribution}

Our models are aimed at starting with a purely gaseous disk, embedded into an existing DM halo. 
Nonetheless, we are aware that these conditions are truly not fully realistic in a cosmological context under two aspects; the existing old stellar component and the isolation of the model DGs, respectively. While cosmological simulations yet lack of sufficient spatial resolutions and physical reliability to provide conditions for low-mass systems at various redshifts, also their temporal mass growth and the mass fraction of their constituents are vaguely known. This allows us to include the stellar mass within the dominating amount of the DM.

For the initial conditions of a stable disk we apply the prescription by \cite{VRH12} to calculate an equilibrium configuration of a rotating gas disk, where the steady-state momentum equation for the gas component in a gravitational potential of gas and DM is solved by
\begin{align}
	\frac{1}{\rho_g}\,\nabla P + (\mathbf v\cdot \nabla) \, \mathbf v = \mathbf{g}_g + \mathbf{g}_h  .
\end{align}
$\rho_g$ is the gas density, $P$ the pressure, $\mathbf v$ the three dimensional velocity and $\mathbf{g}_g$ and $\mathbf{g}_h$ the gravitational acceleration of the gas and halo, respectively. The azimuthal velocity is given here by $v_\Phi = \alpha\,v_c$, where $\alpha$ is the spin parameter and $v_c$ the circular velocity. 
A spin parameter of $\alpha = 0.1$ results in a roundish DG, whereas $\alpha = 0.9$ leads to a flared disk.
Most simulations of isolated galaxies neglect self-gravity when building the initial configuration \citep{VRH12, VRH15}, so that our approach is an advancement compared to commonly used methods.

\subsubsection{The Dark Matter halo}

For the DM halo we assume a spherical isothermal density distribution,
\begin{align}
	\rho_{DM}(r) = \frac{\rho_0}{1 + (r/r_0)^2}
\end{align}
where $\rho_0$ is the central DM density and $r_0$ is the scale radius of the DM halo. For a given halo mass of $M_{DM} = 10^{10}\,\solarmass$ the scale radius, central DM density and the virial radius of the DM halo can be calculated by,  
	\begin{align}
		r_0 &= 0.89 \times 10^{-5} \left( \frac{ M_{DM}}{\solarmass} \right)^{1/2} H_0^{1/2} \,\text{kpc}\\
		\rho_0 &= 6.3 \times 10^{10} \left( \frac{ M_{DM}}{\solarmass} \right)^{-1/3} H_0^{-1/3} \,\solarmass\,\text{kpc}^{-3}\\
		R_{vir} &= 0.016 \left( \frac{ M_{DM}}{\solarmass} \right)^{1/3} H_0^{-2/3} \,\text{kpc}
	\end{align}
where $H_0$ is the Hubble constant and set here to $H_0 = 76\,\text{km}\,\text{s}^{-1}\,\text{Mpc}^{-1}$ for comparison reasons with \cite{VRH12}.

\subsubsection{Two Initial Mass Models}

For the setup of DG models we start from a DM halo mass of $M_{DM} = 10^{10}\,\solarmass$ and distribute the gas mass in rotational equilibrium according to two different spin parameters $\alpha$ of 0.1 and 0.9. The radius $R_{gal}$ of the DG is then defined as the radial distance in the equatorial plane to which the gas density falls to $10^{-27}$ g cm$^{-3}$. This criterion leads to two clearly distinct models: Due to its slow rotation the $\alpha = 0.1$ model forms a spheroidal shape so that the critical radius $R_{gal}$ reaches to 1.3 kpc with a gas mass included of $M_{g} = 6.3\times10^{6}\,\solarmass$ and achieving a maximum rotational velocity of $v_{rot}$ = 2.1 km s$^{-1}$ only. The flatter $\alpha = 0.9$ model, on the other hand, extends to $R_{gal}$ = 9.5 kpc and yields a gas mass of $M_{g} = 1.4\times10^{8}\,\solarmass$ with a maximum rotational velocity of $v_{rot}$ = 30 km s$^{-1}$. 
For regions outside of $R_{gal}$ and densities below $10^{-27}$ g cm$^{-3}$ the gas density is set to $10^{-30}$ g cm$^{-3}$ and a temperature of $T = 10^6$ K to achieve pressure equilibrium with the galaxy. 
According to the DG's size the DM mass within the DG amounts to $M_{DM,\,in\,DG} = 1.2\times10^8\,\solarmass$ for $\alpha = 0.1$ and $M_{DM,\,in\,DG} = 8.4\times10^8\,\solarmass$ for $\alpha = 0.9$. 

To summarise, the parametrization of the shape, i.e. of the angular momentum, physically speaking, leads to two models distinguishable by their different masses. This anti-correlation of axial ratio and galactic mass agrees with the findings by \cite{Sanchez19}.

%%%%%%%%%%%%%%%%%%%%%%%%%%%%%%%%%%%%%%%%
%%%%%%%%%%%%%%%%%%%%%%%%%%%%%%%%%%%%%%%%

\section{The Numerical Setup}\label{sec:numsetup}

To ensure that cooling and SF starts in a steady state, we allow the DG to relax during the first 200 Myr. After that, cooling and SF are switched on. 
The computational box size is chosen to exceed the DG size significantly in order to prevent that x- and y-boundaries affect the DG dynamics and to allow the onstream from the upwind boundary to become detached from a possible reflecting shock. The spatial resolution amounts to 50 pc what means that the scaleheight of the gas disks is resolved by 3 grid cells.

%%%%%%%%%%%%%%%%%%%%
\subsection{Cooling}

For the radiative gas cooling two cooling functions are used, one for temperatures above $10^4\,\text{K}$ \citep*{BH89} and one below $10^4\,\text{K}$ as a combination of the functions by \citet{DM72} and \citet{SKK09}. For the high-temperature regime between $10^4 < T\, \big[\text{K}\big] < 10^8$ \citet{BH89} calculated the cooling functions for hot tenuous plasmas in collisional ionisation equilibrium for temperatures. They focus on the ten most abundant elements (H, He, C, N, O, Ne, Mg, Si, S, Fe) in the universe. Above a temperature of $T =  7\times 10^7\;\text{K}$, cooling comes mainly from the free-free radiation of H and He (bremsstrahlung). In addition to these ten elements, also Ca is tracked in the cdFLASH code. In the low-temperature regime, the fractional ionisation plays an important role, as given by
	\begin{align}\label{eq:IonFrac}
		f_i = \frac{n_e}{n_H}
	\end{align}
with $n_e$ and $n_H$ as the electron and hydrogen number density, respectively. For the low-temperature cooling C, N, O, Si, S, Ne and Fe are taken into account. The electron collisions for these elements are given by a series of equation \citep[Eq. 4 - 12]{SKK09}. The total cooling function is the sum of the cooling functions of these elements.
For simplicity, we start all models with an initial metallicity of 1/3 solar and to account for the evolutionary state of the DGs with solar abundance ratios, both reasonable conditions for dIrrs.

%%%%%%%%%%%%%%%%%%%%
\subsection{Star Formation}

When stars are formed in a grid cell they are traced as stellar particles representing single stellar populations (SSP). The SF is fully self-regulated and follows the description by \cite{KTH95} with a rate of
	\begin{align} \label{eq:SBF}
		\psi(\rho,T) = C_n\, \rho^n\, f(T)
	\end{align}
which contains a power-law dependence on the gas density $\rho$ and on an efficiency function of the temperature $f(T)$ with an efficiency factor $C_n$. The values for $n=2$, $C_2 = 2.575\times 10^8\,\text{cm}^3\,\text{g}^{-1}\,\text{s}^{-1}$ result self-consistently.
$f(T)$ is an exponential function of the temperature $T$ and describes the fraction of the gas that is in the form of molecular clouds,
	\begin{align} \label{eq:SBF_eff}
		f(T) = e^{-T/T_s}
	\end{align}
with a constant $T_s$ = 1000 K. This approach does not require any global scaling relation and was successfully applied to massive spirals by \cite{SHT97} and to dwarf irregular galaxies by \cite{HR02} and \cite{PHR14} leading to realistic SFRs and chemical abundances. In parallel studies we compare this self-regulation recipe with that of density and temperature thresholds as described e.g. by \cite{SSK06} and Kuehtreiber et al. (in prep.) and, secondly, with respect to the initial mass function (IMF) at low SFRs \citep[Steyrleithner et al. in prep.]{HSR17}.  

Since Eq. \ref{eq:SBF}  allows SF even at very low rates, which would increase the number of stellar particles and produce even insignificantly small star clusters, we insert a SFR threshold \citep{PHR14}.
Setting the minimum stellar cluster mass $M_{cl,min}$ that can be formed within a cluster-formation time $\tau_{cl}$ of typically 1 Myr to $100\,\solarmass$, fixes the SFR threshold to $10^{-4} \,\solarmass\,\text{yr}^{-1}$.
Such a short and reasonable cluster-formation time ensures a quick onset of the stellar feedback that is important for SF self-regulation. 
For $\tau_{cl}$ longer than 1 Myr, the stellar feedback is delayed accordingly until $\tau_{cl}$. 
In reality, however, massive stars should have immediately started their feedback. This does not mean that after $\tau_{cl}$ new star particles cannot form in the same grid cell, but the SF criteria might have changed and do not allow further SF due to stellar feedback by the former cluster.

%%%%%%%%%%%%%%%%%%%%
\subsection{Stellar Initial Mass Function}

Once a SSP is formed, it will immediately produce feedback, in the form of stellar radiation and SNeII, depending on the mass of the SSP. By the IMF one can calculate the number of stars for each mass bin. The IMF can be interpreted as a probability function that describes by which probability a star of mass $m$ lies in the interval $[m,\, m+dm]$. It can be described by a power-law
	\begin{align}\label{eq:imf1}
		\xi(m) = k\,m^{-\alpha}
	\end{align}
where $k$ is a normalisation constant and $\alpha = 2.35$ as originally derived for a single power-law by \cite{Salpeter55}. Here we use the multi-section power-law by \cite{Kroupa01} as:
\begin{align}
		\alpha = \left\{
			\begin{array}{l l l}
				0.3 & \dots & 0.01\le m/\solarmass < 0.08\\
				1.3 & \dots & 0.08\le m/\solarmass < 0.5\\
				2.3 & \dots & 0.5\hspace{2mm} \le  m/\solarmass < 100
			\end{array} 
          \right.
\end{align}
The number of stars $N(m)$ and the mass of the cluster $M_{cl}$ can be calculated by
	\begin{align}
		N(m) &= k_1\,\int\limits_{m_{min}}^{m_{max}} \xi(m)\, = 1 \label{eq:imf3.1} \\ 
		M_{cl} &= k_2\,\int\limits_{m_{min}}^{m_{max}} m\,\xi(m)\, dm\,. \label{eq:imf3.2}
	\end{align}
The $k's$ serve as normalisation coefficients and are not the same.
Although the IMF is a probability function, $k_1$ should guarantee that each mass bin N(m) is populated with at least one star at SF. 
For low SFRs this necessary condition can, however, not be fulfilled.
As shown by \cite{PRH15} the IMF can hardly be fulfilled for star clusters of masses below $10^4 \,\solarmass$ so that the stellar feedback is affected at low SFRs \citep[see e.g. Steyrleithner et al., in prep.]{HSR17}.
Since the presented simulations are focussed on the RPS effect, for simplicity the full IMF approach is applied though being aware of the weakness.
Interestingly, for a 1 Myr cluster-formation time this coincides with $10^{-2}\,\solarmass\,\text{yr}^{-1}$ at which the SFR indicators UV and H$\alpha$ begin to diverge \citep{Lee09}.

%%%%%%%%%%%%%%%%%%%%
\subsection{Stellar Feedback}

In our simulations we consider feedback from different sources like by Lyman continuum radiation from massive stars, by type Ia and II supernovae, and by AGB stars.

\subsubsection{Stellar Radiation}

We further improve the SF self-regulation recipe taking into account stellar Lyman continuum radiation by massive stars with masses of $m_\ast > 8\,\solarmass$. 
We calculate the Str\"omgren sphere radius $R_S$ of an HII region. Within the Str\"omgren sphere SF is prevented by the ionisation equilibrium of Lyman continuum radiation vs. the recombination rate of hydrogen in the surrounding ISM by
\begin{align}\label{eq:Rs}
    R_S = \left( \frac{3\,S_\ast}{4\pi\,{n_H}^2\,\beta_2} \right)^{1/3},
\end{align}
$S_\ast$ is the ionizing photon flux, $n_H$ the hydrogen number density, and $\beta_2$ the recombination coefficient within the Str\"omgren sphere. Depending on the resolution, the Str\"omgren radius can be smaller than a typical grid cell, therefore a better sub-grid description is needed, where the mass of the HII region is calculated and subtracted from the cooler gas mass in the cell. As a reasonable temperature of a mean Str\"omgren sphere we set it to $2\times 10^4\,\text{K}$. The gas temperature in a grid cell is then derived as the mass-weighted average of the Str\"omgren sphere and the actual temperature of the remaining gas.

\subsubsection{Type II Supernovae}

Stars with masses larger than $8\,\solarmass$ end their lives as core-collaps SNeII. The feedback is purely thermal, but the SN remnant expands. 
The lifetime $\tau_m$ of a star is related to its mass by the function \citep{RCK09}:
\begin{align}\label{eq:life}
	\tau_m &= \left\{
				\begin{array}{lr}
					1.2\,m^{-1.85} + 0.003\,\text{Gyr} & \text{if}\,m\ge 6.6\,\solarmass\\
					10^{f(m)}\,\text{Gyr} & \text{if}\,m < 6.6\,\solarmass
				\end{array}
			\right.
\end{align}
where
\begin{align}
	f(m) &= \frac{0.334 - \sqrt{1.79 - 0.2232\,\big(7.764 - \log(m)\big)}}{0.1116}\,.
\end{align}
A slight metal dependence of lifetimes is neglected in our case close to solar abundance. 
SNeII eject an energy of $10^{51}\,\text{erg}$, of which 5\,\% are transferred into the ISM. 
The frequently applied and often cited fraction of 10\,\% is unrealistically high for SNeII, because it is derived from a simple 1D hydro model of a single SN explosion \citep{Thornton98} while in reality SNeII explode in star clusters and accumulate to superbubbles. Galaxy simulations with multiple SNeII by \cite{RH06} only derived a value around 2\,\% for SNeII, whereas it is larger for the single SNeIa explosions, see e.g. \cite{Hensler10}.

\subsubsection{Type Ia Supernovae}

A SNIa is the resulting explosion of a C-O white dwarf that reaches the Chandrasekhar mass ($1.44\,\solarmass$) after the accretion of mass of its companion in a binary star system. The SNIa rate depends on the mass ratio between the primary and secondary star in a binary system and is given by \citep{RCK09}
	\begin{align}\label{eq:SNIa}
		R_{SNIa}(t) = A \int\limits_{m_{B,min}}^{m_{B,max}} \int\limits_{\mu_{min}}^{\mu_{max}} &f(\mu)\,\psi(t - \tau_{m_2})\, \nonumber \\
		&\times \xi\big(m_{B},\psi(t - \tau_{m2})\big)\,d\mu\,dm_B \, ,
	\end{align} 
where $A = 0.09$ is a normalisation constant, $m_B$ is the binary mass, $\mu$ is the ratio between the mass of the secondary star $m_2$ and the binary mass, with $\mu_{max} = 0.5$, $\tau_{m_2}$ is the lifetime of the secondary star and $\xi$ is the IMF (sect. 2.5.). 
The distribution function for mass ratios in binary systems is given by
	\begin{align}
		f(\mu) = 2^{1+\gamma}\, (1+\gamma)\, \mu^\gamma
	\end{align}
\citep{MR01}, where the parameter $\gamma = 2$ is based on statistical analyses that show mass ratios of one more likely. We implied this recipe to the cdFLASH code \citep{PRH15}. However, nowadays smaller $\gamma$ is preferred and set here to $\gamma \simeq 0.5$ (Recchi, priv. commun.). 
With Eq. \ref{eq:SNIa} the number of SNeIa in each mass bin is calculated. If a star ends as a SNIa, all its mass is returned to the ISM.

\subsubsection{Asymptotic Giant Branch Stars}

Stars with masses below $8\,\solarmass$ return mass back to the ISM during the asymptotic giant branch (AGB) phase. A fraction of these stars will end their life as SNIa.

\subsubsection{Chemical Feedback}

At the end of their lives stars not only return mass to the ISM, but also nucleosynthesis products as chemical feedback. SNeII enrich their surroundings mostly with $\alpha$-elements like O, Ne, Mg, Si, S and Ca, whereas SNeIa contribute mostly Fe. AGB stars release mostly N and C to their close surroundings. We use the stellar yields from \citet{MBC96} for stellar masses $m_\ast = (1 - 4)\,\solarmass$, and \citet{PCB98} for masses above $6\,\solarmass$ with a linear interpolation between $4$ and $6\,\solarmass$.

%%%%%%%%%%%%%%%%%%%%%%%%%%%%%%%%
\subsection{Wind Module}

In order to study the effect of RP we implemented a so-called wind module to the cdFLASH code \citep{Mit13,PRH15}. For the motion through the ICM its density varies according to the radial ICM density profile for the Virgo cluster of \cite{VCC01} as
	\begin{align}\label{eq:VirgoDens1}
		\rho_{ICM}(r) = \rho_0 \left( 1 + \frac{r^2}{r_c^2} \right)^{-\frac{3\beta}{2}}
	\end{align}
with $\beta = \frac{1}{2}$ and the core radius $r_c = 13.4\,\text{kpc}$. 
While the central density amounts to $\rho_0 = 10^{-26}\,\text{g}\,\text{cm}^{-3}$ we set $\rho_0$ to $10^{-28}\,\text{g}\,\text{cm}^{-3}$ only, because we do not favour radial infall towards the cluster centre but instead aim at considering a parabolic path that only touches the cluster outskirts and at studying the DG stripping at very low ICM densities.

Since the acceleration of an infalling galaxy by the gravitational potential of the above-mentioned density profile is small or even zero (for $\beta = \frac{2}{3}$), for simplicity we assume a constant relative velocity of the galaxy with respect to the hot ICM.  
For an efficient computational modelling, the DG is set into a wind tunnel where the wind moves towards positive z direction (upwards in Fig. \ref{fig10}). The DG's equatorial plane extends in the x-y-plane, what means that the wind hits the DG face-on. Therefore, Eq. \ref{eq:VirgoDens1} has to be transformed from a radial-dependent to a time-dependent equation, i.e.\ with $r = r_{max} - v\,t$ to
	\begin{align}\label{eq:VirgoDens2}
		\rho(t) = \rho_0 \left( 1 + \frac{(r_{max} - v t)^2}{r_c^2} \right)^{-\frac{3\beta}{2}}
	\end{align}
with $r_{max}$ as the initial distance from the Virgo cluster centre. 
Starting from the density $\rho(r = r_{max}) = 10^{-30}\,\text{g}\,\text{cm}^{-3}$, $r_{max}$ is given from Eq. \ref{eq:VirgoDens2}.
With this close-up simulation we can constrain the computational domain and gain a higher spatial resolution to study galaxy internal effects like the location of SF.

%%%%%%%%%%%%%%%%%%%%%%%%%%%%%%%%
\section{Models}\label{sec:models}

In order to study the RPS effect on DGs we perform simulations of DGs' evolution in isolation with identical parameters as the RPS models. For a better understanding of the mass dependence, we model two different masses, a low-mass DG with $6.3 \times 10^6\,\solarmass$ gas mass and a more massive DG with $1.4 \times 10^8\, \solarmass$. For reasons of numerical expenses old stellar components are comprised in the DM mass that are factors of 19 and 6 larger than the gas masses of the low and the high-mass model, respectively. 
In addition, the shape of the DG as determined by the spin parameter $\alpha$ according to $v_\phi\ = \alpha \ v_{circ}$ is varied to 0.1 (slow rotation) for the low-mass and 0.9, i.e. almost rotationally flattened, for the high-mass DGs. Because we aim at studying also RPS in different environments, i.e. DGs in galaxy groups and clusters, we chose three different wind velocities, 100, 290, and 1000 km s$^{-1}$, what means that the RP varies by steps of an order of magnitude due to its square-dependence on v. 
In order to limit the number of simulations presented here and to get a clearer insight into the RP efficiency, the low-mass models (assigned as rpsLM) are performed with the higher two relative velocities of 290 and 1000 km $s^{-1}$, while the massive DG (named rpsHM) are exposed to the two lower (group) velocities of 100 and 290 km $s^{-1}$. In order to distinguish the RPS effect we compare the models with the evolution of isolated DGs without any circumgalactic gas also H$\alpha$ to avoid gas infall: isoGT2 for the low-mass DG and isoHM2 for the massive one.  

The simulations are then started at rest and without any physical process acting except gravitation for 200 Myr to relax the system before the wind, respectively, the relative motion of the model galaxy is switched on. 
In these models, the velocity of the galaxy is kept constant, while the ICM density changes with time, according to the density profile given by Eq. \ref{eq:VirgoDens2}. The maximal ICM density and RP are reached at different simulation times depending on the infall velocity (see table \ref{tab:runs}). After the maximal ICM density is reached, it is kept constant. This simplification is reasonable, because we do not intend to trace the real path within the cluster and because with decreasing density the RPS will also weaken. 
The minimum (initial) and maximum (after 3 Gyrs) RP for the models amount to $8.41\times 10^{-16}$ dyne cm$^{-2}$ and $8.41\times 10^{-14}$  dyne cm$^{-2}$ for the 290 km s$^{-1}$ velocity model, and to $10^{-14}$ dyne cm$^{-2}$ and $10^{-12}$ dyne cm$^{-2}$ for the 1000 km s$^{-1}$ velocity model, respectively.

Table \ref{tab:runs} provides an overview of the simulations with the parameters of the DGs and wind properties such as velocity and ICM density. As mentioned above, all models start with a metallicity of 0.3 $Z_{\odot}$.

\begin{table*}
	\caption{\small{Initial parameters for different simulations. $M_g$ is the initial gas mass, $M_{DM}$ is the DM mass within the galaxy, $\alpha$ is the spin parameter which controls the rotational velocity, with its maximal velocity $v_{rot}$. $R_{gal}$ is the radius of the DG, defined in section \ref{subsec:IC}, $v_{wind}$ the velocity of the ICM wind and $\rho_{0,ICM}$ the maximal ICM density. }}
	\centering
		\begin{tabular}[c]{lccccccc}
			                \hline
			runID & $M_{g}$ & $M_{DM}$ & $\alpha$ & $v_{rot}$ & $R_{gal}$ & $v_{wind}$ & $\rho_{0,ICM}$ \\
				  & $[\solarmass]$ & $[10^8\,\solarmass]$& & $[\text{km}\,\text{s}^{-1}]$& $[\text{kpc}]$ & $[\text{km}\,\text{s}^{-1}]$& $[\text{g}\,\text{cm}^{-3}]$\\
			\hline
			{\bf isoGT2}   & $6.3\times 10^6$ & $1.2$ & 0.1 &  2.1 & 1.3 & -         & -  \\
			{\bf isoHM2}  & $1.4\times 10^8$ & $8.4$ & 0.9 &  30.0      & 9.5 & -          & - \\
			{\bf rpsLM2}   & $6.3\times 10^6$ & $1.2$ & 0.1 &  2.1 & 1.3 & 290   & $10^{-28}$ \\
			{\bf rpsLM4}   & $6.3\times 10^6$ & $1.2$ & 0.1 &  2.1 & 1.3 & 1000 & $10^{-28}$ \\
			{\bf rpsHM1}  & $1.4\times 10^8$ & $8.4$ & 0.9 &  30.0       & 9.5 & 290 & $10^{-28}$ \\
			{\bf rpsHM2}  & $1.4\times 10^8$ & $8.4$ & 0.9 &  30.0       & 9.5 & 100 & $10^{-28}$\\
			\hline
		\end{tabular}
	\label{tab:runs}
\end{table*}

%%%%%%%%%%%%%%%%%%%%%%%%%%%%%%%%
\section{Results}\label{sec:results}

%%%%%%%
\subsection{Comparison between low-mass and high-mass Dwarf Galaxies}

In order to learn about the effect of RP on the evolution of DGs, among others e.g. about the SF, we compare low-mass and high-mass DGs within a wind environment with their isolated counterparts. For this, we vary the relative wind speed and once also its density (see Table 1).

%%%%%%%
\subsubsection{rpsLM2 vs isoGT2}

\begin{figure}
   \centerline{ \includegraphics[width=\columnwidth]{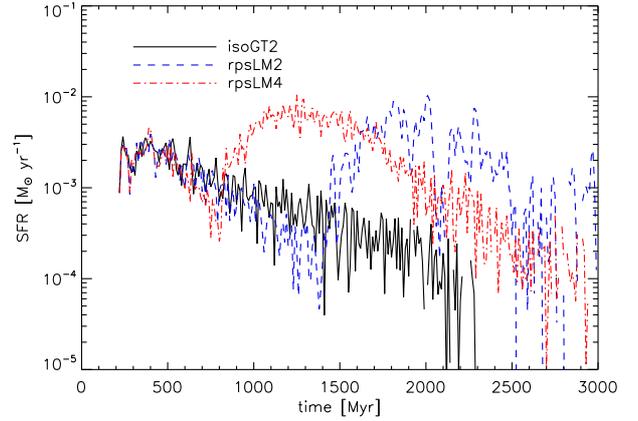} }
   \caption{\small{Star formation rate in 10 Myr bins for the LM  models, the isolated DG isoGT2 (solid black) and the DGs within a wind environment of slow (rpsLM2, dashed blue) and fast (rpsLM4, dash-dotted red) speed. }}
   \label{fig1}
\end{figure}

At first, we compare the low-mass DG model within a wind environment (rpsLM2) with the isolated DG (isoGT2), both simulations starting with the same initial conditions. After the wind is initialised the RP is increasing and one would expect the SFR to be (slowly) increasing due to gas compression or at least to be constant over time. In contrast, also starvation can be expected due to the gas loss by stripping.  

\begin{figure}
   \centerline{ \includegraphics[width=\columnwidth]{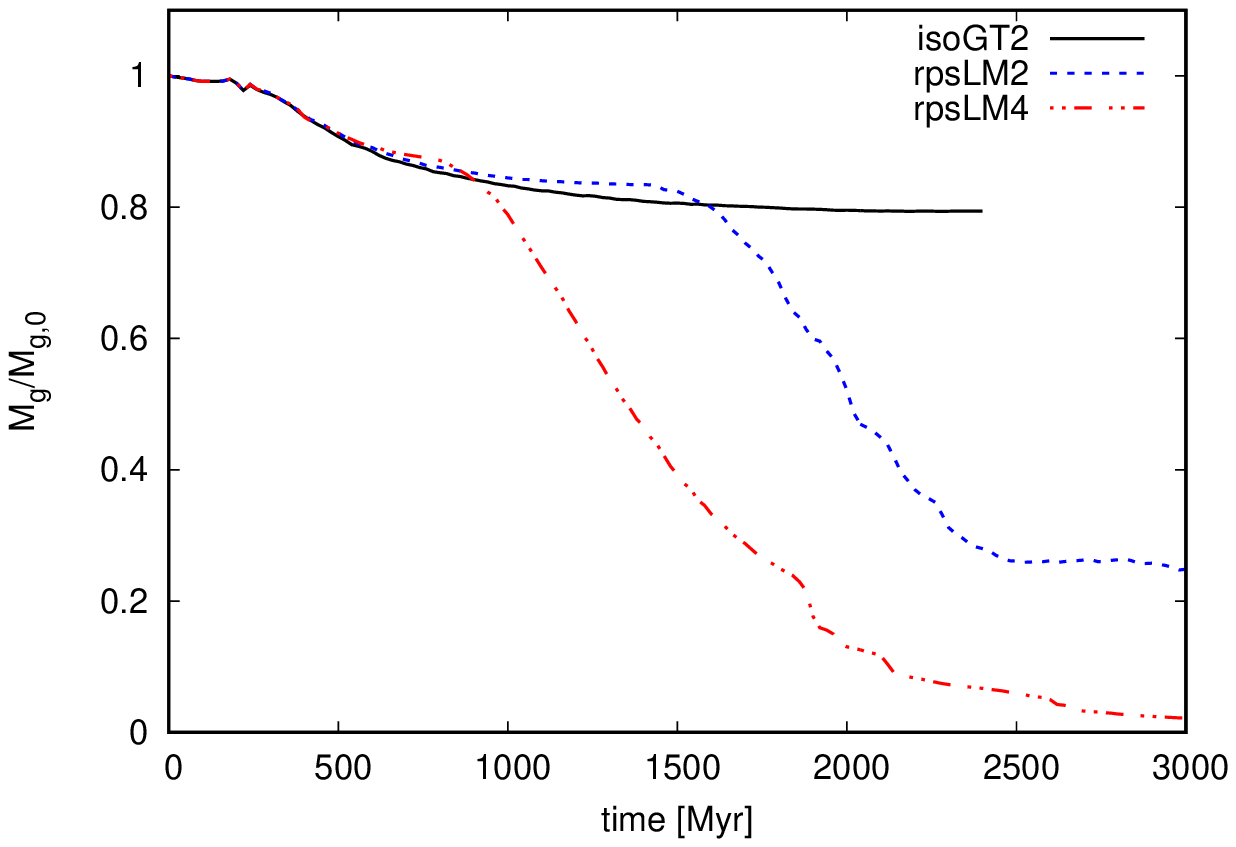} }
   \centerline{ \includegraphics[width=\columnwidth]{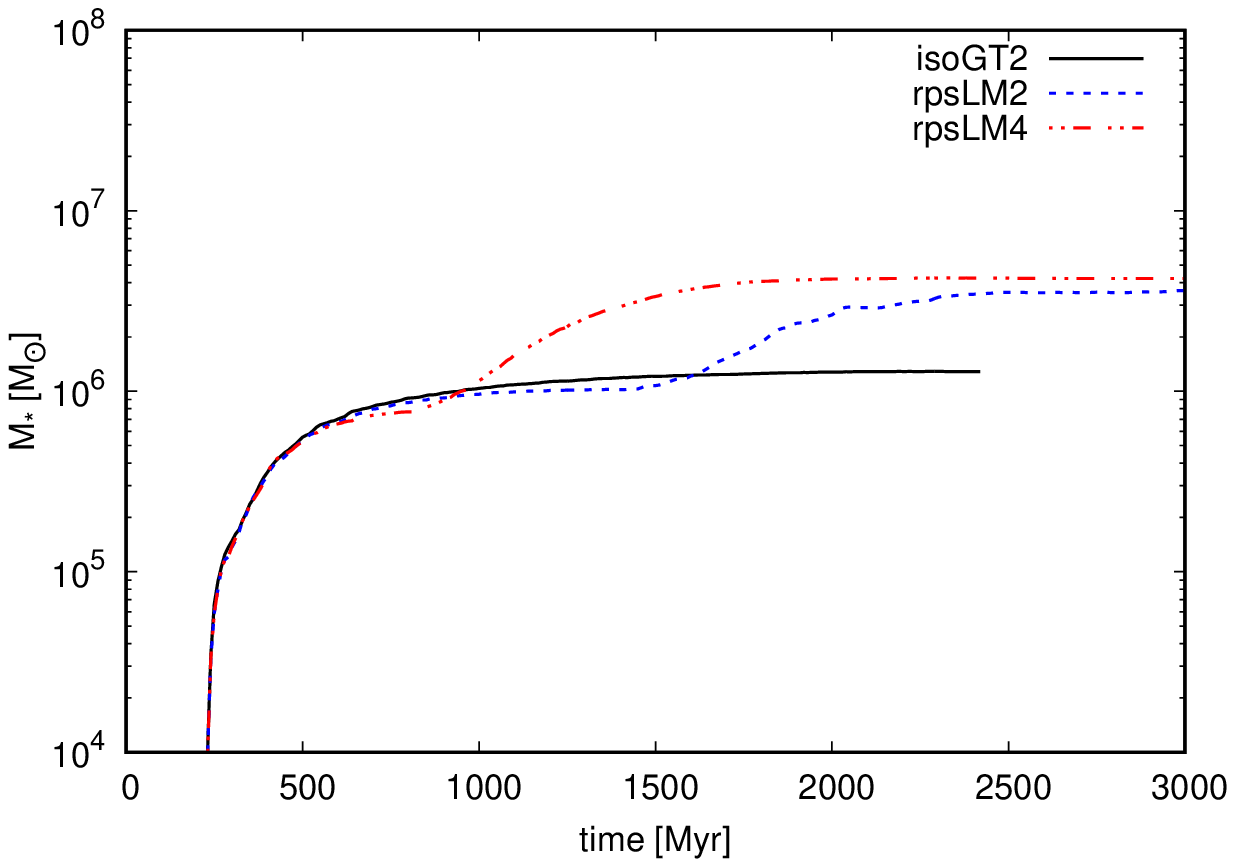} }
   \caption{\small{Gas mass normalised to the initial gas mass (top) and absolute stellar mass (bottom) for the isolated DG isoGT2 (solid black) and the DGs within a wind environment, rpsLM2 (dashed blue) and rpsLM4 (dashed-dotted red). }}
   \label{fig2}
\end{figure}

The simulations show an unexpected behaviour. During the first epoch of  $1.5\,\Gyr$ the SFR of the rpsLM2 model is decreasing more quickly than for isoGT2 (Fig. \ref{fig1}), meaning that the wind suppresses the SF. To quantify this effect, a linear fit is performed for both SFRs for a simulation time of $500\le t_{sim}\,[\Myr] \le 1500\,\Myr$\footnote{The time was chosen to be in this interval to ensure that only the effect of the wind on the SFR will be analysed.}. 
The resulting slope of the SFR for rpsLM2 with $( 1.056 \pm 0.096 ) \times 10^{-3}\,\solarmass\,\text{yr}^{-1}\,\Myr^{-1}$ is steeper than for isoGT2 given by $(8.05 \pm 0.75)\times 10^{-4}\,\solarmass\,\text{yr}^{-1}\,\Myr^{-1}$. The steeper decline of the SFR at RP can be understood by the perturbance and the reduction of the DG's outermost ISM. Nonetheless, Fig. \ref{fig2} reveals that obviously the lower SFR is mainly caused by RP, because the gas mass remains higher and the stellar mass lower than that of the isolated model between 1.0 and 1.5 Gyr. Not before the ICM wind reaches its maximum density at $t_{sim} = 1500\,\Myr$, the RP gets strong enough to compress the DG gas sufficiently and, by this, triggers SF. This leads to a rise of the SFR of rpsLM2 by almost two orders of magnitude within 100 $\Myr$ reaching $10^{-2} \solarmass\,\text{yr}^{-1}$ again for almost 1 Gyr with a subsequent slow decline. Afterwards the SFR begins to strongly oscillate as it becomes also visible for the isolated model. The reason is probably the varying amount of cool gas at small gas fraction.   

The effect of the RP on the SFR is also visible in the gas and stellar masses in Fig. \ref{fig2}. The lower SFR in model rpsLM2 at the beginning leads to a slightly higher gas mass. Although the stellar mass is accordingly slightly lower at the beginning than that of the isolated DG, it rises due to the triggered SF and even exceeds the isolated model. 

\begin{figure}
   \centerline{ \includegraphics[width=\columnwidth]{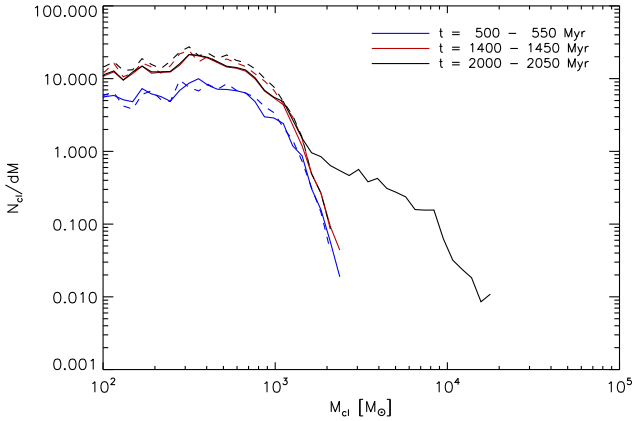} }
   \caption{\small{The cluster mass function in $50\,\Myr$ age bins for the runs isoGT2 (dashed lines) and rpsLM2 (solid lines) shows the number of clusters per mass bin (dM) as a function of the cluster mass.}}
   \label{fig3}
\end{figure}

For the runs isoGT2 and rpsLM2 Fig. \ref{fig3} shows the cluster mass function (CMF) at different simulation times in age bins of $50\,\Myr$. Due to the higher SFR the cluster mass (CM) extends by one order of magnitude (Fig. \ref{fig3} black solid line). 
The huge amount of gas loss of $75\,\%$ (Fig. \ref{fig2}, top) is not directly due to the mass consumption by SF, but can be explained by the increase of the released SNII energy by a factor of 10. 
The higher energy release due to the larger number of more massive stars in each cluster facilitates large gas clouds to be pushed off from the galaxy and carried away by the ICM wind, because the gravitational potential is not strong enough to keep them bound. 
The turbulences caused by the energetic feedback sometimes clear the galaxy's centre of the gas, resulting in gaps in the SFR.

%%%%%%%%%%%%%%}
\subsubsection{Slow (rpsLM2) vs. fast (rpsLM4) low-mass DG models}

As one can discern from Fig.\ref{fig1} for the fastest model rpsLM4 investigated as a proxy of cluster infall, reasonably, the SF trigger by RP sets in earlier by approximately 600 Myrs than for rpsLM2. While SF in rpsLM4 also happens in stripped-off clouds outside the main galaxy body, in model rpsLM2 the SF continues within this DG though with decreasing intensity. 
Although the difference in the infall velocity is large, both models reach the same maximum SFR of up to $10^{-2}\, \solarmass\, \text{yr}^{-1}$. 
Furthermore, the model rpsLM4 has lost almost all of its gas after $3\,\text{Gyr}$ (see top of Fig. \ref{fig2}) mainly due to the earlier RP maximum compared to rpsLM2. 
The differences in the gas properties can also be seen by comparing the initial and final gas density at the centres of these models. At first, both have an initial gas density of $\rho_i = 5.9 \times 10^{-25}\,\text{g}\,\text{cm}^{-3}$, rpsLM2 exhibits a final gas density of $\rho_f = 1.3 \times 10^{-25}\,\text{g}\,\text{cm}^{-3}$. 
While the gas density for rpsLM2 decreases by a factor of 4.5, rpsLM4 loses all its gas.
Despite the difference in the gas-loss rate, gas density and the SFR, both models reach the same stellar mass (Fig. \ref{fig2}, bottom) but one can speculate that rpsLM2 will continue its SF though at an extremely low rate and concentrated to the centre only.

%%%%%%%
\subsubsection{Massive DGs with ram pressure (rpsHM1) vs. its isolated pendant isoHM2}

\begin{figure}
   \centerline{ \includegraphics[width=\columnwidth]{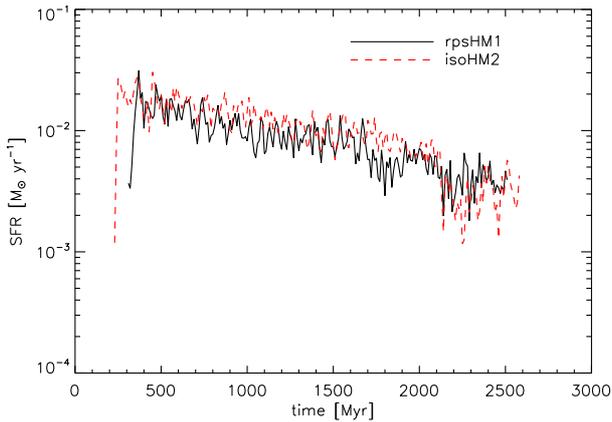} }
   \caption{\small{Star-formation rate in 10 Myr bins for the isolated DG (isoHM2) and the DG within a wind environment (rpsHM1), respectively.}}
   \label{fig4}
\end{figure}

As further analysis we compare the more massive DGs, the rpsHM1 model with the isolated isoHM2. For DGs with higher masses the SFR is not that strongly affected by the wind (Fig. \ref{fig4}) as the LM models and gets slightly suppressed. 

A linear fit to the SFR with time shows that they are in fact not that different, because the slope of the SFR for rpsHM1 of $( 2.92 \pm 0.15 )\times 10^{-4}\, \solarmass\, \text{yr}^{-1}\, \Myr^{-1}$ is comparable to the slope of isoHM2 with $( 3.63 \pm 0.17 )\times 10^{-4}\,\solarmass\,\text{yr}^{-1}\, \Myr^{-1}$. 
Compared to low-mass models, where the gas loss is facilitated because the gas is pushed out by the high SNII energies, the gas of high-mass DGs is purely lost by RP, because the deeper potential well hampers the formation of large-scale outflows. 

\begin{figure}
   \centerline{ \includegraphics[width=\columnwidth]{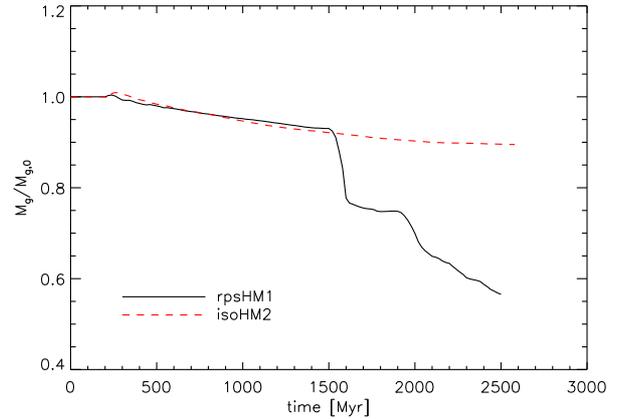} }
   \caption{\small{Gas mass normalised to the initial gas mass for the isolated DG (isoHM2) and the DG within a wind environment (rpsHM1), respectively.}}
   \label{fig5}
\end{figure}

Fig. \ref{fig5} shows the normalised gas mass of the high-mass DGs. 
When the RP becomes strong enough near its maximum and must have exceeded a critical pressure after a simulation time of $t_{sim} = 1.5\,\Gyr$, where $P_{ram} \ge P_{therm} \simeq 3\times 10^{-14}\, \text{dyne}\, \text{cm}^{-2}$, most of the galaxy's gas is pushed away instantaneously and, therefore, the gas mass drops steeply but by $\sim 20\%$ only. Coincidently the SFR is slightly declining. After the instantaneous stripping of the warm tenuous gas, its removal by Kelvin-Helmholtz instability continues. 
The gas loss is not only fuelled from the galaxy's outskirts by RPS, but also reduces the central gas density by a factor of 16, only from an initial density of $\rho_i = 1.6\times 10^{-24} \,\text{g}\,\text{cm}^{-3}$ to a final density of $\rho_f = 10^{-25} \,\text{g}\,\text{cm}^{-3}$.

Finally, the slow HM DG model rpsHM2 is not extensively presented and discussed here, because the RP amounts to not more than 10$^{-14}$ dyne cm$^{-1}$, i.e. less than $P_{therm}$. Reasonably, the DG evolution is hardly affected by the RP and only small displacements of gas and SF from the gaseous disk (see Fig.\ref{fig7}) without significant oscillations are insufficient for gas stripping and heating of the stellar velocity dispersion (see next sect.).

%%%%%%%
\subsection{Where does the enhanced SF occur?}

\subsubsection{The Ram-pressure perturbed Disk}

\begin{figure}
   \centerline{ \includegraphics[width=\columnwidth]{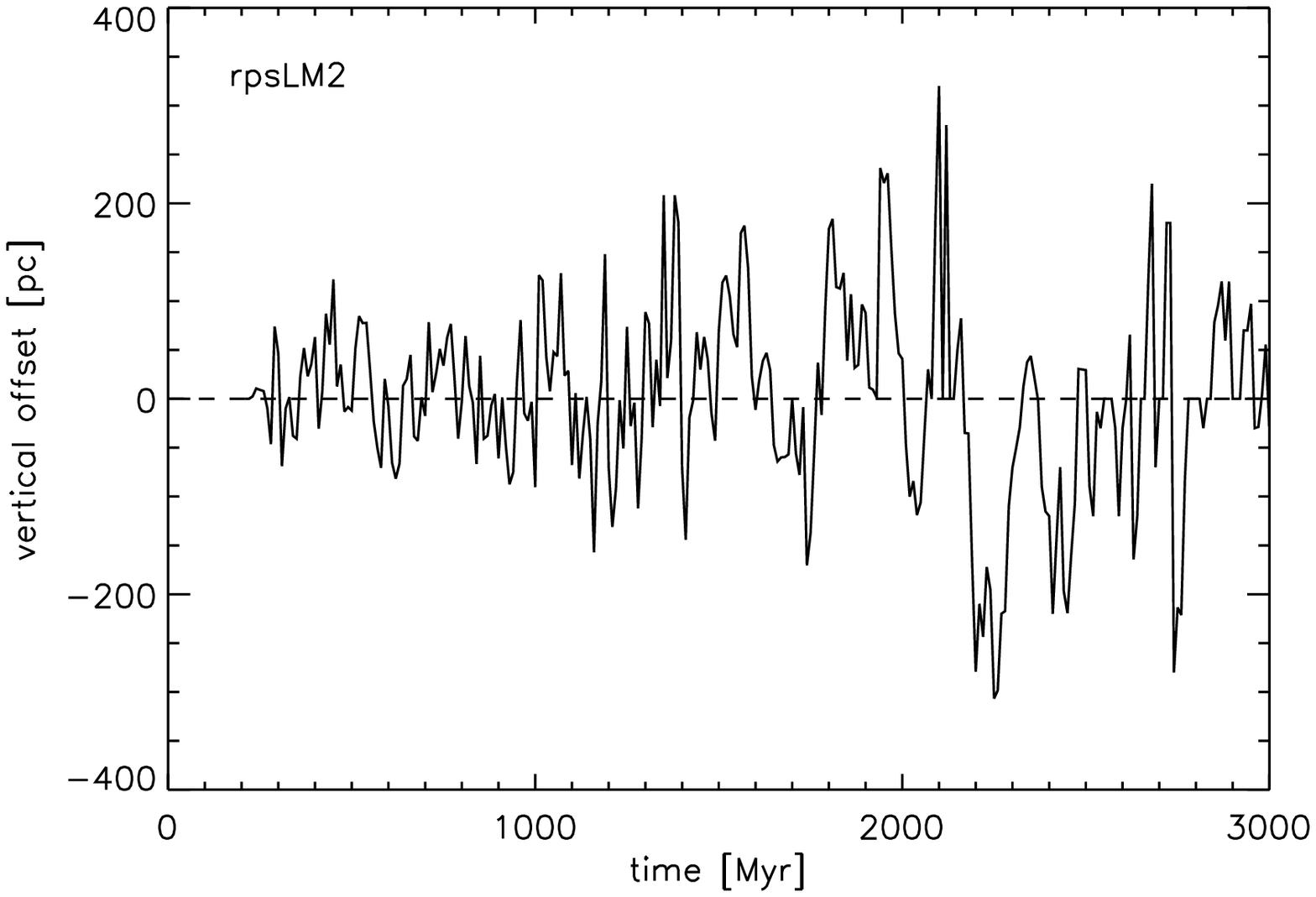} }
   \centerline{ \includegraphics[width=\columnwidth]{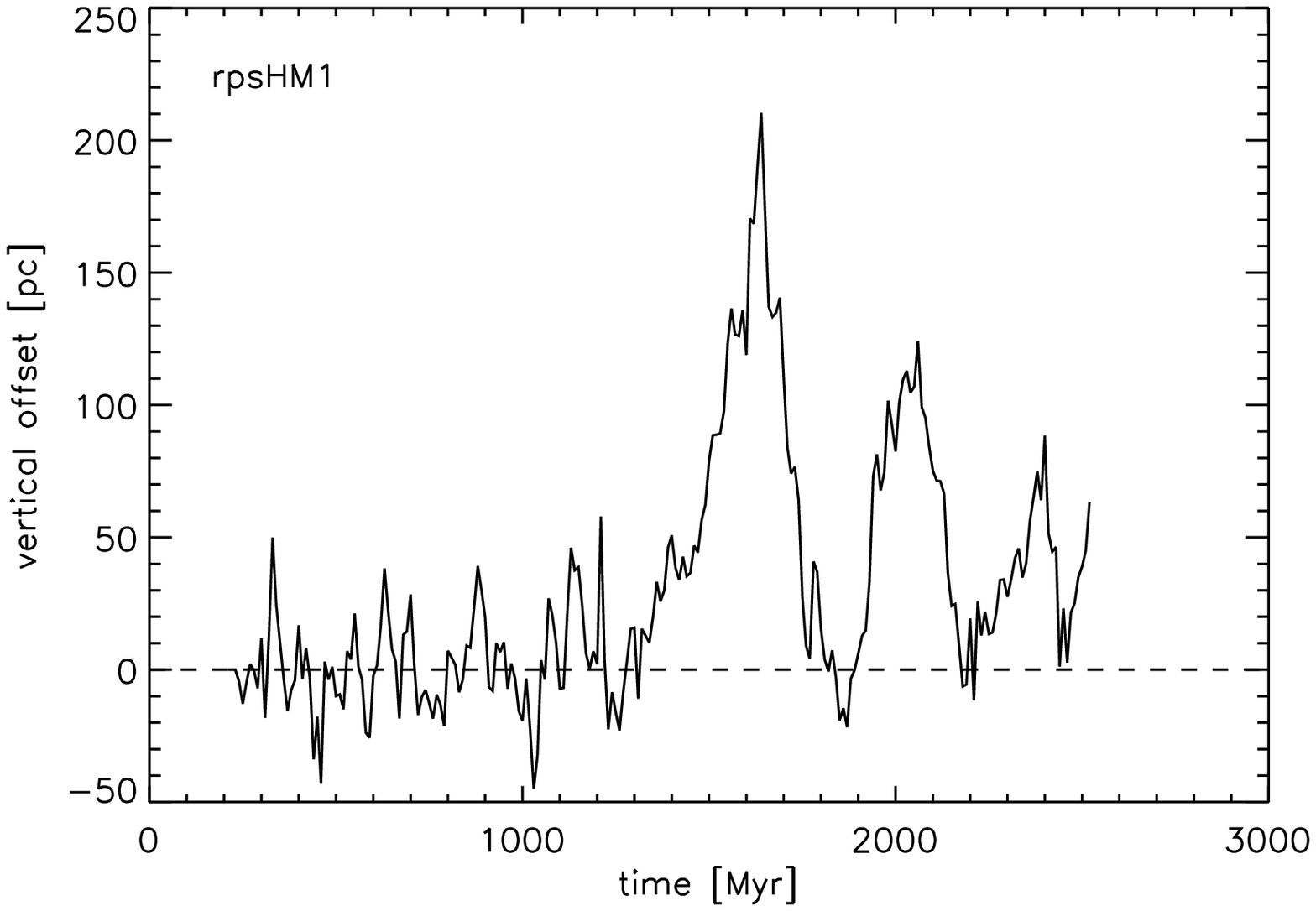} }
   \caption{\small{Vertical offset of the centre of SF for the run rpsLM2 (top panel) and rpsHM1 (bottom panel).}}
   \label{fig6}
\end{figure}

To quantify the effect of RP on the SF, we derive the shift of the mass-weighted centre of SF in z-direction. If the SF is assumed to happen in RP stripped clouds, which are pushed by the wind, then the centre must be shifted towards positive z-directions. 
The centre of SF is calculated by 
\begin{align}
	z_\psi = \frac{1}{\psi^{tot}} \sum\limits_{i=1}^n \psi_i\,z_i
\end{align}
where $\psi_i$ is the SFR, $z_i$ the z-coordinate for each cell, and $\psi^{tot} = \sum_i\psi_i$ the total SFR. When the wind hits the galaxy face-on (which is the case for all simulations in this paper), then the wind compresses the gas at the windward side of the galaxy. Since the ICM density is increasing over time as well as the RP, then the gas compression increases too, which should lead to a higher SF at the windward side. 
Since the galactic centre of mass is located at the origin of the coordinate system, the centre of higher SF caused by the RP compression at the DG front should lie towards negative z direction. Fig. \ref{fig6} shows this quantitatively for the run rpsLM2 (top panel) and rpsHM1 (bottom panel). 
For both runs the wind velocity of $v_{wind}$ = 290 km s$^{-1}$ and the RP are the same. The RP reaches its maximum at $t_{sim} \simeq 1.4\, \text{Gyr}$. 

At first, the centre of SF is oscillating around the zero line (dashed line), i.e. the equatorial plane, but when the RP is strongest, the centre of SF as well as the gas are displaced towards the positive z-axis. 
This behaviour is opposite to that one expected. It is not so prominently discernible for the model rpsLM2, because gas is more rapidly lost into the z-direction and, therefore, reaches slightly larger offsets, but for a short while only. On the other hand, the high-mass model rpsHM1 experiences a strong and extended push at about 1.5 Gyr when the RP approaches its maximum and 20 \% of the gas is lost (see Fig. \ref{fig5}), namely, when after the first strong SF onset further wide displacements follow to smaller heights because the gas is hardly lost. 

One should keep in mind, that these offsets are relatively small compared to the size of the entire simulation box of $\pm 12.8\, \text{kpc}$ and to the smallest grid cell of $50\, \text{pc}$, resolving the scaleheight by three cells. By comparing the extent of the displacement with the vertical size of the galaxy, the perception changes. The LM model has (initial) vertical extent of $1.2\, \text{kpc}$ due to the roundish shape, whereas the HM model reach $0.5\,\text{kpc}$. 
To remind the reader, the outer edge of a model is defined where the ISM density has dropped to the value of $10^{-27}\,\text{g}\,\text{cm}^{-3}$. 
For the model rpsLM2 this means that the displacement happens between one and four grid cells, whereas for rpsLM4 the displacement is increasing with time up to 1 kpc, which is at the outer edge of the galaxy. For the model rpsHM1 the displacement is between one and six cells, and for rpsHM2 always between one and two cells around the equatorial plane.

\begin{figure}
   \centerline{ \includegraphics[width=\columnwidth]{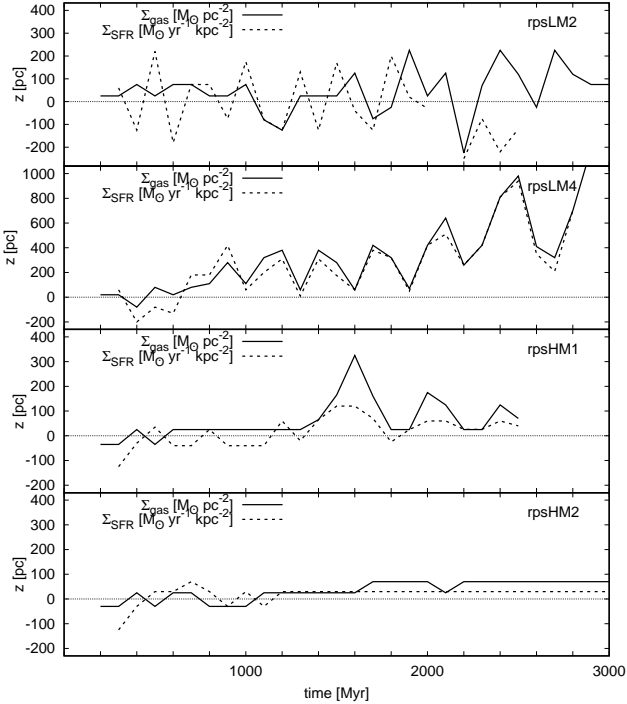} }
   \caption{
\small{Maximum values of the gas and SFR surface density integrated over the x- and y-axis, respectively, every 100 Myr for the simulations rpsLM2, rpsLM4, rpsHM1 and rpsHM2 (from top to bottom). The solid line represents the gas surface density, the dashed line the SFR surface density and the dotted line indicates the coordinate centre of the simulation box. }}
   \label{fig7}
\end{figure}

The centre of SF seen in Fig. \ref{fig6} does not show why it is shifted to the positive z-axis, i.e. down-stream. Is the expectation wrong, that the SF should occur where the gas gets compressed, or is the whole galaxy pushed back along the wind direction in such a way that the centre of the galaxy's baryonic matter does not coincide anymore with the centre of mass and thus the coordinate centre?
To get a more quantitative insight into the effect of the RP on the entire galaxy, Fig. \ref{fig7} shows the maximum of the gas surface density and the SFR surface density at various z integrated over the xy-planes, respectively. 
There clearly exist difference in the response of the gas to the RP between the LM and HM galaxies, as also on the position of the SF. This is due to the different shapes of the galaxies. The flat extended disk of the HM models are more exposed to RP due to the fast rotation.  

Comparing the maxima of the gas and SFR surface densities of the LM DGs, respectively, reveals different behaviours. In rpsLM2 both maxima are oscillating about $z=0$, and while those of the SFR are mostly further displaced from the equatorial plane than the gas they happen mostly also before the gas maxima. In contrast, for rpsLM4 both components oscillate coincidently, but are shifted to larger and temporarily increasing distances. Due to the stronger RP it happens to the back-side of the DG only. 
While the oscillations in rpsLM2 are mainly due to stochastic effects, for rpsLM4 they are supported by the RP which pushes the gas off from the galaxy towards the wind direction. 

The HM DG models also differ in the RP strength due to their different velocities.
This effect is visible at the beginning, after the wind was initialised. For a higher wind velocity of 290 km s$^{-1}$ of rpsHM1 the initial gas compression is stronger so that the maximum of the SF surface density is shifted to the negative z-axis, i.e. towards the windward side of the galaxy. For comparison, this is less pronounced for rpsHM2 at a wind velocity of v = 100 km s$^{-1}$ only. The effect of the stronger RP for rpsHM1 is also more prominent at later times, where the gas and SF  oscillate. For both galaxies, rpsHM1 and rpsHM2, the SF surface density is shifted to the upwind side, indicated by the fact that the maximum of the SF lies beneath that of the gas. 

\begin{figure}
   \centerline{ \includegraphics[width=\columnwidth]{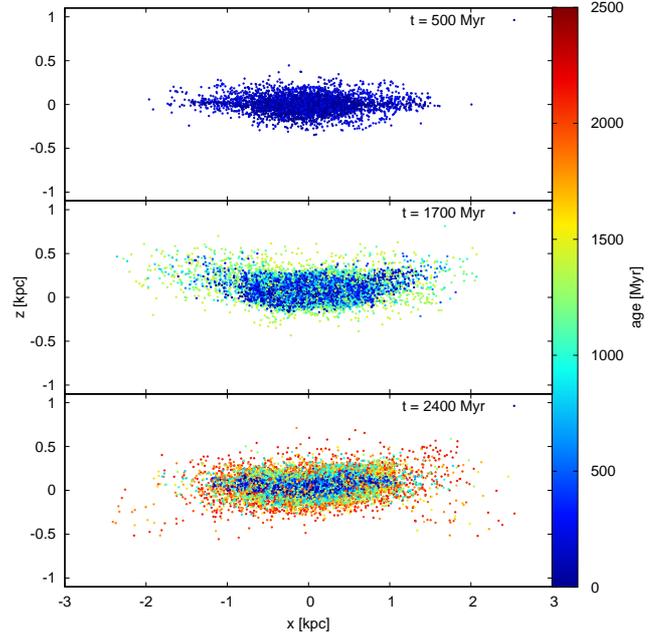} }
   \caption{\small{Distribution of star clusters formed during the simulation of model rpsHM1 at three different snapshots. The cluster ages are coloured from blue (youngest) to red (oldest), since the simulation started. Star clusters which are formed from ram-pressure displaced gas clouds but are still bound by the DG have to oscillate about the equatorial plane and transform the younger stellar population into an ellipsoidal shape.}}
   \label{fig8}
\end{figure}

\begin{figure}
   \centerline{ \includegraphics[width=\columnwidth]{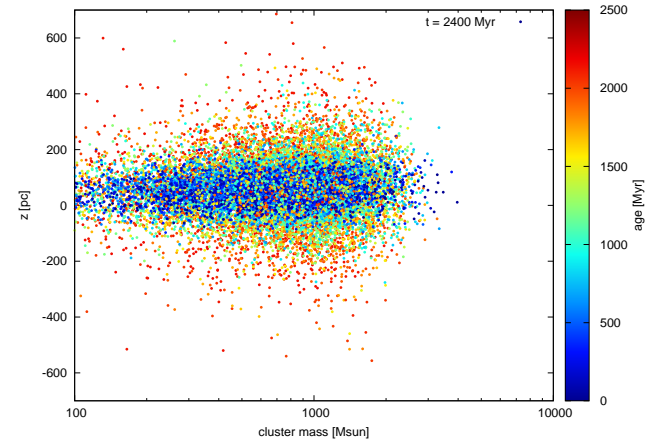} }
   \caption{\small{Cluster mass versus the vertical position of formed star clusters. The cluster ages are coloured from blue (youngest) to red (oldest), since the simulation started, for the model rpsHM1 at a simulation time of 2400 Myr.}}
   \label{fig9}
\end{figure}

Finally, one has to emphasise two main effects of the location of SF sites and its consequences: RP displaces the SF maxima, for the LM DGs towards the front of the gas maxima, for HM DGs mainly ''behind'' the gas and clearly ''above'' the equatorial plane. Since these  star clusters formed in HM DGs are still bound to the DG, they have to oscillate about the equatorial plane (see Fig. \ref{fig8}). By this, the young stellar population becomes ellipsoidal and more extended than the old stars so that a radial and vertical colour gradient should arise of old stars more concentrated towards disk and centre and formed long before the cluster infall than the younger (bluer) stars. Since those latter are formed at larger vertical height (see Fig. \ref{fig9} for better visibility) and oscillate through the equatorial plane, they are also discernible with higher velocity dispersion, making the brightness shape more elliptical. 
While the ongoing SF sites are constrained to smaller radial ranges due to RP gas loss, also the vertical extent shrinks again, so that the colour gradients become uncertain, this even more at very low SFRs. 
From Fig. \ref{fig9} it becomes visible that only low-mass star clusters are formed so that the completeness of the IMF is doubtful and, by this, whether the evolution of the colour gradient can be understood by SSPs.

%%%%%%%
\subsubsection{Star Formation in Ram-Pressure Stripped Clouds}

When the RP is strong enough, a large amount of gas will be stripped off the galaxy. Either entire gas clouds will be pushed out, or stripped gas condenses to blobs that will be further carried away. Then it must be further examined whether these blobs remain gravitational bound and fulfil the SF criteria (sect. 3.2.). 

Most of such blobs have a relatively high gas temperature and low density, both not appropriate for SF. They are e.g. heated up by friction with the hot ICM and turbulences. Eventually most of the blobs start to cool again and could then form stars. To take this into account, the total gas mass of each blob is calculated, but only for those cells that have temperatures below $2\times10^4$ K, and compare them with the corresponding Jeans mass\footnote{To calculate the Jeans mass, the average gas density and temperature, respectively, of each blob are used.} $M_J$,
\begin{align}
	M_J = \sqrt{\frac{6}{\pi}} \left( \frac{k\,T}{G\,\mu} \right)^{3/2} \rho^{-1/2}
\end{align}
where $T$ represents the average temperature, and $\rho$ the average gas density of each blob, $G$ is the gravitational constant, $k$ the Boltzmann constant, and $\mu$ the mean molecular weight here set to $\frac{5}{3}$ . The results are listed in table \ref{tab:blobdata2}. If the total gas mass of a blob is larger than its $M_J$, the blob should collapse and form stars. 
The comparison of gas and Jeans masses in table \ref{tab:blobdata2} directly shows that the rpsLM2 blobs do not reach M$_J$, while all blobs of model rpsHM1 do. Surprisingly, SF occurs in none of them. Only the fast rpsLM4 model produces stripped-off clumps (B3 and B4) which are conditioned for SF with the extreme blob B4. Even though a factor of 5-8 less massive than the HM1 gas clouds B5 and B6, B4 is capable to fulfil the SF criteria for eight times the SF timescale of 1 Myr. The physical cause is obviously the strong compression to much denser clumps than the other gas blobs.

\begin{table*}
	\caption{Mean gas density $\overline\rho$ and temperature $\overline T$, respectively, as well as the total gas mass $M_{g,tot}$ of each blob and the calculated Jeans mass $M_J$ at the onset of star formation for comparison. The most right column gives the number of computing grid cells ($x \times y \times z$) of the maximum expansion of the blobs in each coordinate (for its explanation, see subsection \ref{subsec:sfblobs}). The properties are measured after the blobs got unbound from the DG and reached a distance of more than 2 kpc downstream of the galaxy. 
}
	\centering
		\begin{tabular}[c]{llccccc}
			                \hline
runID & Blob & $\overline\rho$ & $\overline T$ & $M_{g,tot}$ & $M_J$ & no. of \\ 
           & no.  & $[10^{-26}\,\text{g}\,\text{cm}^{-3}]$ & $[\text{K}]$ & $[10^5\solarmass]$ & $[10^5\solarmass]$ & grid cells \\ 
			\hline
rpsLM2 & B1 & $4.2$ & $10639.9$ & $0.05$ & $3.2$ &$4 \times 8 \times 6$ \\
rpsLM2 & B2 & $4.3$ & $\,\,\,9348.9$ & $1.2$ & $2.6$ &$10 \times 9 \times 16$ \\
rpsLM4 & B3 & $64$  & $\,\,\,6989.2$ & $1.8$ & $0.4$ &$5 \times 6 \times 15$ \\
rpsLM4 & B4 & $48$  & $\,\,\,9355.5$ & $1.4$ & $0.8$ &$7 \times 8 \times 9$ \\
rpsHM1 & B5 & $3.5$ & $10196.5$ & $9.1$ & $3.3$ &$30 \times 30 \times 70$ \\
rpsHM1 & B6 & $3.1$ & $10203.1$ & $11$  & $3.5$ &$17 \times 60 \times 30$ \\
rpsHM1 & B7 & $3.6$ & $10274.1$ & $5.4$ & $3.3$ &$50 \times 25 \times 20$ \\
rpsHM1 & B8 & $4.4$ & $10242.2$ & $3.3$ & $3.0$ &$15 \times 15 \times 20$ \\
rpsHM1 & B9 & $3.6$ & $10270.1$ & $12$  & $3.3$ &$70 \times 25 \times 25$ \\ 
		\hline
		\end{tabular}
	\label{tab:blobdata2}
\end{table*} 

\begin{figure}
   \centerline{ \hspace{2.5cm} \includegraphics[width=3in]{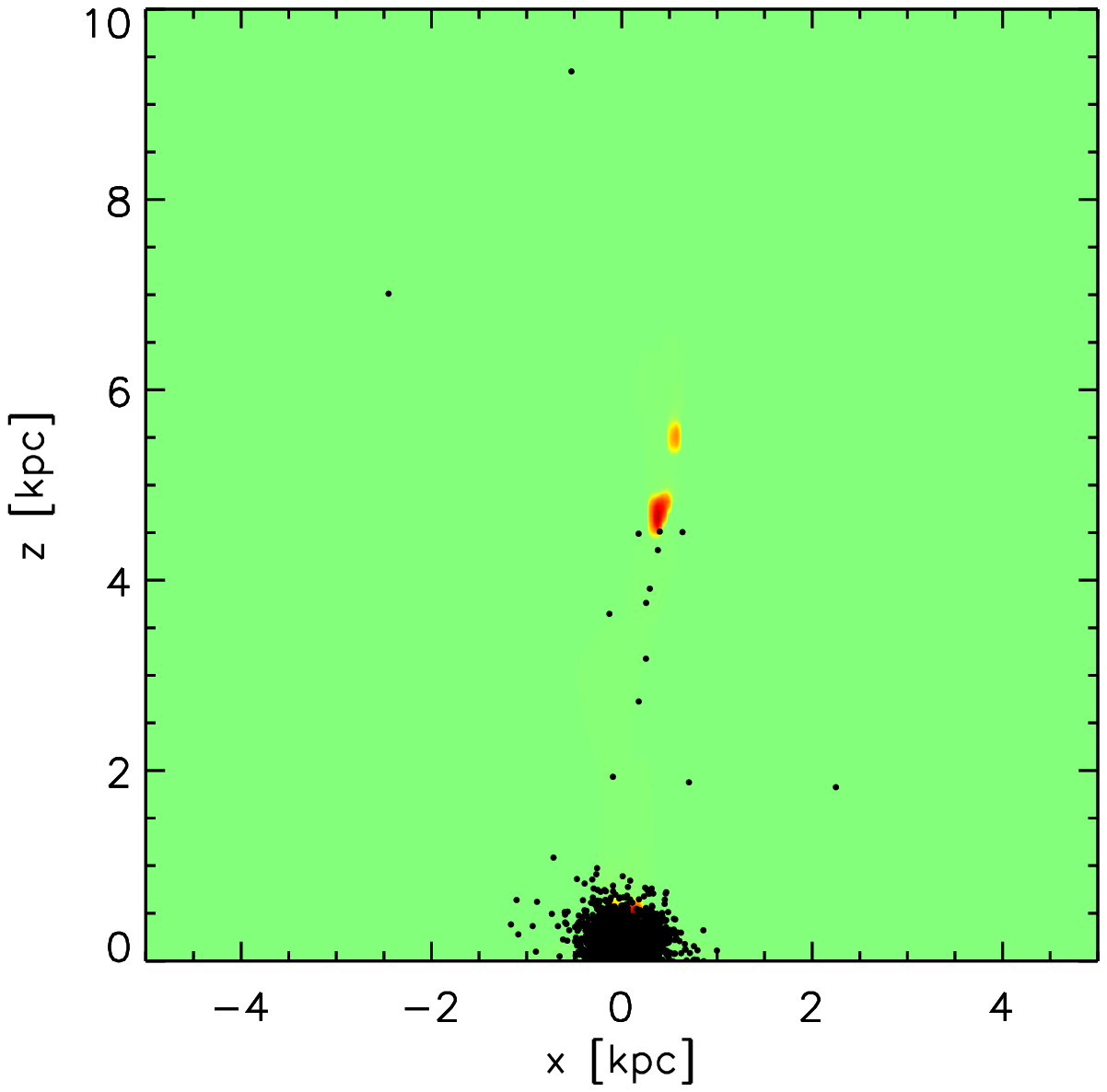} \hspace{-4.35cm} \includegraphics[width=3in]{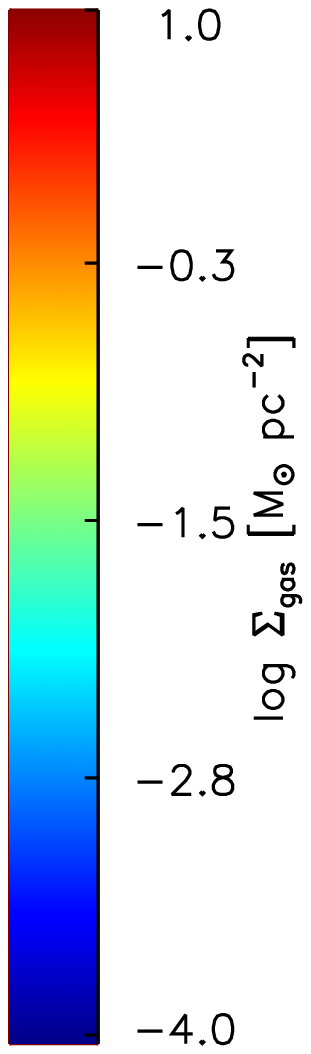} }
   \caption{\small{Snapshot of the gas surface density (filled contours) and particle positions (black dots) in the xz-plane for the run rpsLM4 at $t_{sim}$ = 2300 Myr. Visible at the centre of this snapshot at z = 4.5 kpc is the blob B4, leaving a trail of newly formed star clusters behind. The two star clusters on the left-hand side of the box at a distance of above z = 7 kpc are remaining from blob B3 which already  left the simulation box. }}
   \label{fig10}
\end{figure}

A closer analysis of the simulation rpsLM4 reveals that the actual SF can be identified in at least two of the stripped blobs (B3 and B4), while all blobs of the other models do not (for physical reasons see the consideration in sect. 3.3). Blob B3 has a starburst event not before far away from the host galaxy's centre, at a distance of more than 6\,kpc. Blob B4 exhibits several starburst of SF at a distance of more than 2\,kpc while it is stripped away. Fig. \ref{fig10} shows a snapshot of the run rpsLM4 at a simulation time of $t_{sim}$ = 2300 Myr with blob B4 and the positions of the stellar particles. While blob B4 forms stars, the gaseous component still experiences further acceleration by the RP, but the stars do not. Therefore, blob B4 leaves behind a stream of ''naked'' star clusters. The results are listed in table \ref{tab:blobSF}.

\begin{figure}
   \centerline{ \includegraphics[width=\columnwidth]{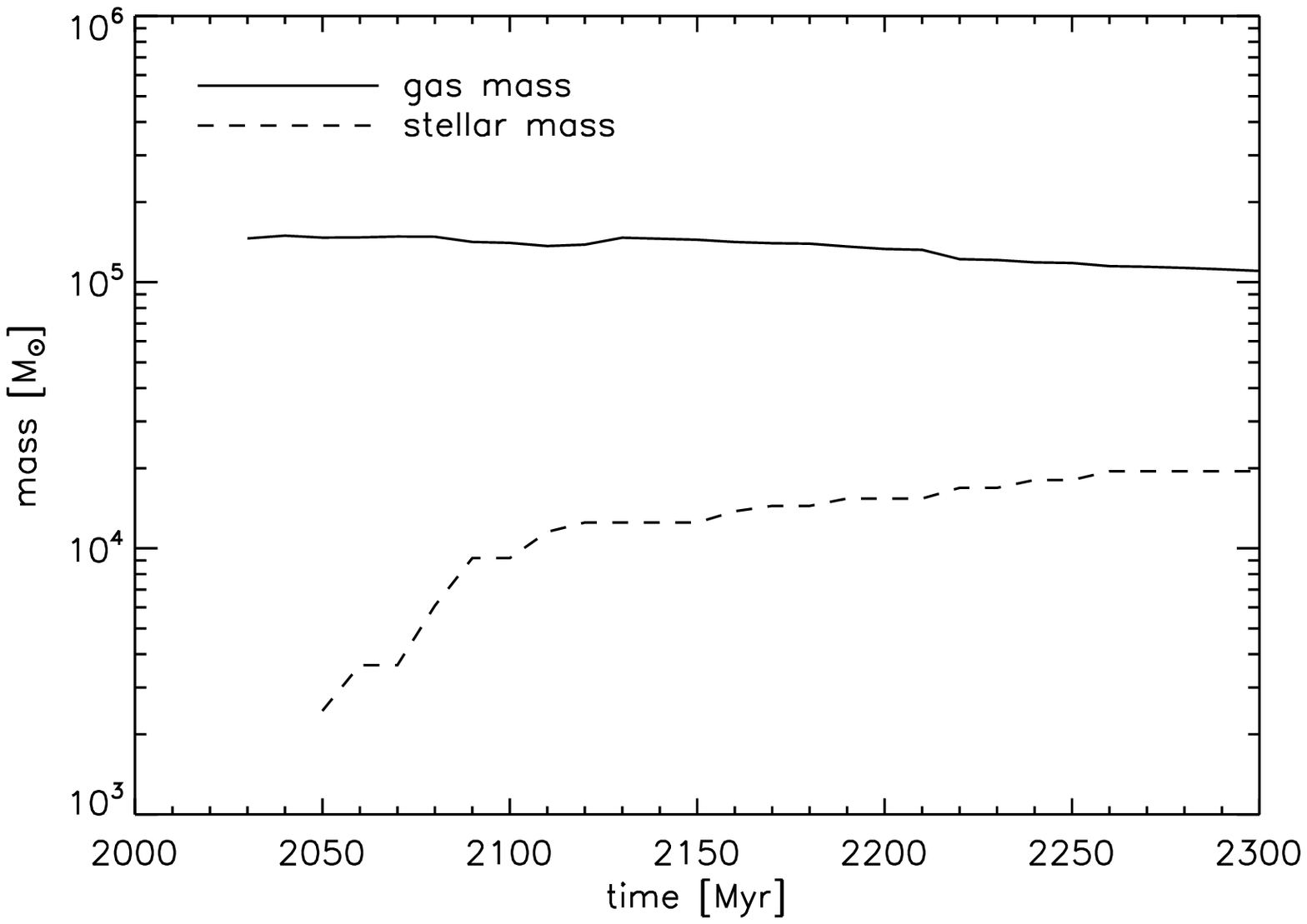} }
   \centerline{ \includegraphics[width=\columnwidth]{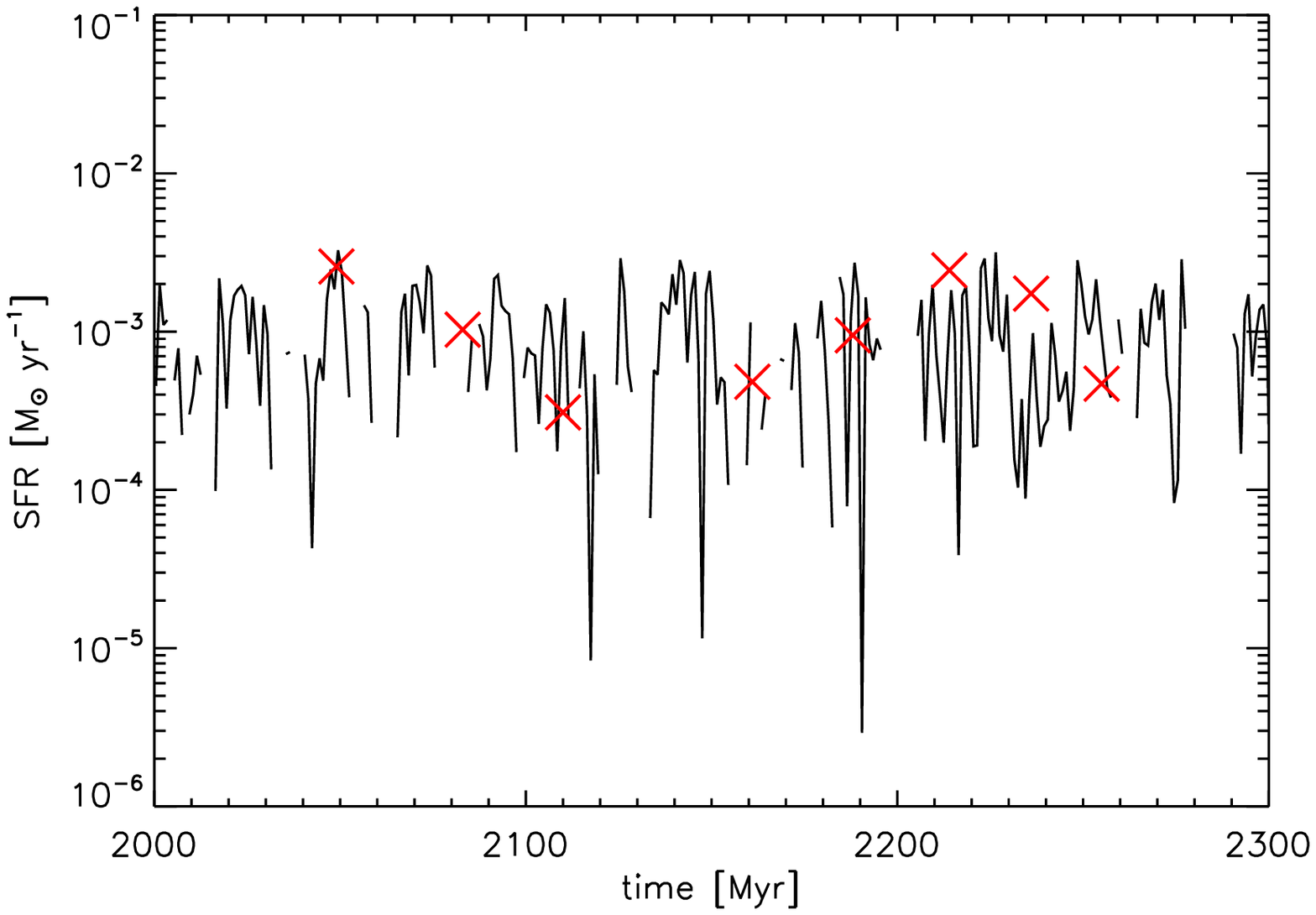} }
   \caption{\small{Top: Gas and stellar mass, respectively, for the blob B4, which exhibits a continuous SF over $\sim 200\,\text{Myr}$; Bottom: Global SFR (solid line) of rpsLM4 during the starburst of blob B4. The peaks of the starburst for blob B4 are shown as red x-symbols. }}
   \label{fig11}
\end{figure}

Interestingly, these blobs experience a very short but relatively strong starburst with SFRs of the order of $10^{-3}\,\solarmass\,\text{yr}^{-1}$. The top panel of Fig. \ref{fig11} shows the gas and stellar mass, respectively, for the blob B4, which exhibits a continuous SF for $\sim 200\,\text{Myr}$. 
One has to be aware that the gas is further accelerated while the star clusters keep their velocity when they were formed. Therefore, the gas leaves 'naked' star clusters behind which cannot appear as HII regions.
The bottom panel of Fig. \ref{fig11} shows the galactic SFR in 1-Myr bins during the epoch of starbursts in blob B4, which are shown as red x-symbols for the eight SF events. When their values reach the galactic SFR or even exceed this, the SF is preferably happening in the stripped-off blobs.

\begin{table}
	\caption{Number of star-forming  particles $N_{part}$, total stellar mass $M_\ast$, duration of starburst $\Delta t$ and the corresponding SFR for the star-forming blobs.}
	\centering
		\begin{tabular}[c]{lccccc}
			                \hline
Blob & $N_{part}$ & $M_\ast$ & $\Delta t$ & SFR \\ 
             &  & $[\solarmass]$ & $[\text{Myr}]$ & $[\solarmass\,\text{yr}^{-1}]$ \\
			\hline
	B3 & 2 & 2392 & 1 & $2.4\times10^{-3}$ \\
	B4 & 3 & 2536 & 1 & $2.6\times10^{-3}$ \\
	B4 & 3 & 3110 & 3 & $1.0\times10^{-3}$ \\
	B4 & 6 & 3096 & 10 & $3.1\times10^{-4}$ \\
	B4 & 2 & 1935 & 4 & $4.8\times10^{-4}$ \\
	B4 & 1 & 948 & 1 & $9.5\times10^{-4}$ \\
	B4 & 2 & 2445 & 1 & $2.4\times10^{-3}$ \\
	B4 & 2 & 1737 & 1 & $1.7\times10^{-3}$ \\
	B4 & 3 & 1408 & 3 & $4.7\times10^{-4}$ \\
			\hline
		\end{tabular}
	\label{tab:blobSF}
\end{table}

%%%%%%%%%%%%%%%%%%%%%%%%%%%%%%%%
\section{Discussion}\label{sec:discussion}

As we can demonstrate by our presented models, RP can produce four main effects on gas-rich DGs at their cluster infall, which we discuss in detail in the following subsections.
\\
The \textit{first} and obvious one is the evacuation of gas by means of the drag force exerted by hot gas. \\
As a \textit{second} effect, but not generally acting, the galactic SF is enhanced by compression of the ISM. Surprisingly, for the rpsHM DG the SFR decreases equally as in the isolated model, whereas for the LM models the SFR drops steeper until the RP exceeds a critical value (in rpsLM2 of $P_{ram} = 8.41\times10^{-14}$ dyne cm$^{-2}$ according to Eq. \ref{eg:Pram}). Then the SFR suddenly rises and peaks almost two orders of magnitude higher, but falls again due to the complete or at least partial gas evacuation. \\
The \textit{third} relevant, but hardly detectable effect we discovered in the rpsHM1 model is the thickening of the stellar disk by newly formed stars in ISM clouds which are swept to the back of the disk plane. 
Since these newly formed star clusters are gravitationally bound and fall back towards the galactic disk plane and oscillate about it, by this, thickening the disk. \\
\textit{Finally}, SF can occur in stripped-off clouds, but in the models though with the same environmental conditions only eventually as it is also observed. In particular, the circumstances for this dichotomy must be explored in detail with respect to the physical conditions.

\subsection{Gas loss by Ram-pressure Stripping of Dwarf Galaxies}

Although the Gunn-Gott formula \citep{GG72} is a simple approach to quantify the stripping radius and, by this, the gas loss of the galactic disks by means of RP perpendicular to the disk surface, observations of the "real" stripping effect and numerical simulations find a surprisingly well agreement with the formula. Because of its simplicity, the formula cannot be fully applied  to inclined on-stream and the subsequent stripping phase due disk re-expansion and Kelvin-Helmholtz instability as disentangled by \cite{RH05}. Further complication for the stripping effect arises due to the ISM inhomogeneities in the sense that the momentum transfer by the RP depends on the density contrast between the hot dilute ICM and the up to orders of magnitude denser galactic ISM. Observations reveal that DGs in galaxy clusters for which theory requires their gas to be completely evacuated by RPS, do not fulfil this expectation and retain gas in their centres. 
Our simulations provide a physically reasonable explanation for this phenomenon: The compression of the ISM due to the RP leads to less sensitivity to RP due to its enhanced density.

\subsection{Star-formation Enhancement due to Ram Pressure} \label{subsec:sfrp}

That gas-free dEs in the Virgo cluster have suffered a last strong SF epoch before the gas was pushed out and the SF ceases, is also reproduced by the simulations. While the rpsHM models experience almost the same total SFR as isolated DGs with a reasonable decrease due to gas consumption, their star-forming regions emerge behind the equatorial following the RP shifted gas. Since their maxima lie closer to the plane than the gas, the SF happens in front of the gas maxima. 

The rpsLM simulations show two surprising different effects. While the SFR drops steeper than for the isolated model due to the fast RPS gas loss, the stronger compression of the remaining galactic gas at the onstreaming side triggers the SF to increase steeply by two orders of magnitude. The SF maxima are also displaced from the equatorial plane but even further off than the gas maxima and their spatial oscillations not always coincide with the gas.

\begin{figure}
   \centerline{ \includegraphics[width=\columnwidth]{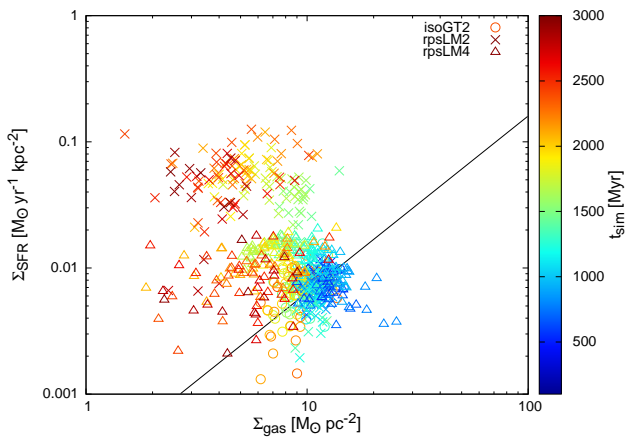} }
   \caption{\small{The models isoGT2 (circles), rpsLM2 (x-symbols) and rpsLM4 (triangles) at various simulation times with respect to the Kennicutt-Schmidt relation (black solid line). (For discussion see subsection \ref{subsec:sfrp}) }}
   \label{fig12}
\end{figure}

Fig. \ref{fig12} shows the models isoGT2, rpsLM2 and rpsLM4 compared to the Kennicutt-Schmidt (KS) relation. Reasonably, the isolated DG stays around the KS line, whereas the DGs within a wind environment experience an up to two orders of magnitude higher SF triggered by the strong RP. They are shifted to the region of starburst DGs, whereby the low-speed RP of rpsLM2 (see Fig. \ref{fig12} red x-symbols) exerts a stronger trigger efficiency than the fast rpsLM4 model.

\subsection{Dwarf Galaxy Transformation and Stellar Population Gradients}\label{subsec:trans}

In both models, LM and HM respectively, star clusters form at distances up to 200 pc above the equatorial plane and are still gravitational bound so that they fall back and oscillate about the plane, by this, leading to a thicker disk of newly formed stars than the old stellar population. This unexpected effect seems to depend strongly on the properties of galaxies and environment. At first, the galaxy mass should be reasonably low in order to allow a discernible dislocation of the lifted star-forming  clouds. This shift is not as strong for the HM models as preferably for the LM DGs (see Fig. \ref{fig6} and \ref{fig7}). The result is a heating of the stellar velocity dispersion and, by this, a thickening of the stellar distribution. 
On the other hand, the thickening becomes better discernible for the HM models (as rpsHM1 shown in Fig. \ref{fig8} and \ref{fig9}), because the LM models are initially more roundish and the SFR maxima are derived under the inclusion of unbound star clusters (see sect. \ref{subsec:sfblobs}).
This effect leads to a morphological transformation from an elongated disky shape to an ellipsoid. While this shape transformation is generally believed to be achieved by subsequent harassment within the galaxy cluster \citep{Smith15}, our finding allows a new and physically reasonable path to dEs, at least a pre-conditioning of dE shaping.  
Moreover, the relative ICM speed should be in a slow range to exert a cloud lifting only but no push-away. 

As an observable signature that should be studied those cluster dEs should show a negative vertical stellar age gradient and a positive blue colour gradient or at least flat ones, as well as a radially increasing stellar velocity dispersion. This effect might already be indicated by a few dEs in the Coma cluster with radial changes in age \citep{Kol11} and by the negative or flat age gradients derived in a few Virgo cluster dEs \citep{Urich17} from the SMAKCED sample \citep{Janz12}. 

On the other hand, the colours of star clusters depend on their initial mass function. While the HM models experience a SFR above the critical value of 10$^{-2}\, \solarmass$ yr$^{-1}$ and can fill the IMF \citep{PHR14}. The LM models evolve the whole time below this limit, at which the SF indicators of H$_{\alpha}$ and UV diverge, see e.g. \cite{Lee09}. 
This means, on the other hand, that a sufficient number of massive stars can hardly be formed to allow for the assumption of the colour evolution of a regular single-stellar population. This large uncertainty, however, makes predictions of colour gradients and of age deviations as interpretable results of RP very vague.

\subsection{Star-Forming Blobs}\label{subsec:sfblobs}

There are several ways to withdraw gas clouds from a galaxy: At first when the RP is strong enough (for velocities larger than $100\,\text{km}\,\text{s}^{-1}$) to directly push away the material; secondly, close and fast encounters with other galaxies lead to harassment; and finally, enhanced SN feedback caused by a starburst drives out not only hot SN gas but the mass-loaded ISM also. 
Although all simulations in this work\footnote{Except for rpsHM2 which shows a smooth continuous gas removal.} exhibit stripped-away gas clumps, that should be Jeans unstable, only two blobs in one simulation (rpsLM4) experience active SF. While there are a few results in the literature with active SF in simulated ram-pressure stripped clouds (e.g. \cite{KKF08,KSS09,RBO14}), these concern massive spiral galaxies only. 

Here we present for the first time simulations (rpsLM4) of a RPS DG with active SF in stripped clouds and study its dependence on the relative velocity. Comparisons with observations reveal some remarkable similarities. For instance the DG IC3418 in the Virgo cluster, studied by \cite{HSN10,FGS11,JKR13,KGJ14}, is moving through the ICM with a velocity of $\sim 1000$ km s$^{-1}$ (similar to rpsLM4). 
\cite{HSN10} assume a RP of $2.7\times10^{-11}$ dyne cm$^{-2}$, which corresponds to an ICM density of $2.7\times10^{-27}$ g cm$^{-3}$ and is slightly more than one magnitude larger than for model rpsLM4. 
\cite{BC03} show by their simulations that RP can trigger SF in a molecular cloud of spiral galaxies by compressing the molecular gas. They also assume that a cloud can be prevented from collapsing, when it is surrounded by diffuse HI gas. But if this envelope is stripped away, the cloud is directly exposed to the RP, which leads to a compression and subsequent SF.  

The LM model with the fastest wind onstream, on the other hand, experiences SF also in gas clouds that are stripped-off from the galactic body out to distances of 1 kpc and more, at which the clusters are not bound anymore to the galaxy. This allows the formation of freely floating star clusters, as it is observed since recently in the DG VCC\,1217 (aka IC\,3418) \citep{HSN10, FGS11, JKR13, JCC14, KGJ14}, in the low-mass SBc galaxy ESO+137-001 \citep{SDV07}, and e.g. in the edge-on stripped Virgo spiral NGC\,4254 \citep{Bos18}.  
Recently, we detected a new DG in the VESTIGE survey of the Virgo cluster \citep{BFF18} IC 3476, that suffers edge-on RPS (Boselli et al. 2020, in prep.). Star-forming regions are preferably located in the arm which is stripped-off more extendedly due to the rotation direction.
 
Our models show that SF happens in stripped clouds at stronger RP. As a matter of fact, all galaxies with star-forming RPS clouds move with extremely high relative velocities of $1000\,\text{km}\,\text{s}^{-1}$ and above through the ICM. 
For the high-mass models the stripped gas clouds are probably not massive enough due to the lower velocities. Either they are destroyed by evaporation or dynamically or they need more time to cool, collapse and form stars. In the latter case the resulting SF would happen far away in the galaxy's wake, as shown in the simulations of massive galaxies by \cite{KKF08,KSS09,RBO14} and from some observations in stripping action, as e.g. \cite{SDV07,Bos18}. The simulations are carried out with almost the same fast relative velocity. 

The bias of triggered SF in stripped blobs can be understood by the different cloud masses: stronger RP pushes off more massive clouds which are suitable for SF after cooling and approaching rest so that the external ICM pressure has to replace the $M_J$ criterion by a Bonnor-Ebert mass \citep{Bonnor,Ebert}. In contrast, slower RP either lifts up massive clouds only and/or strips off small clouds and more diffuse gas, both not capable for SF. 
Nevertheless, one must keep in mind that most simulations do not include heating processes by the thermal/radiative environment so that the stripped gas cools only and reaches the criteria for SF internally. In particular, numerical particle schemes as Smoothed-particle Hydrodynamics (SPH) recipes with fixed particle masses switch-on SF when the $M_J$ criterion is reached in addition to implied temperature and densities thresholds. 

Although \cite{RH08} have proven that the local RPS criterion in a galaxy, i.e. the local gravity, obeys the same analytical formula as the Rayleigh-Taylor instability (RTI), for individual gas clouds like those stripped off the criterion for their destruction by means of RTI must consider their local self-gravity. To resolve a gas cloud gravitationally, the numerical gravity solver needs to cover at least three cells in each dimension. To ensure that the selected blobs are also gravitationally resolved, their expansion in each dimension $(x \times y \times z)$ is listed in table \ref{tab:blobdata2}. 
From this one can discern that only in 2 out of 9 clouds the Jeans mass $M_J$ is not reached. For the 7 exceeding $M_J$, the lighter ones B3 and B4 are on average denser by more than one order of magnitude than e.g. B5 and B6, although these are more massive by a factor of 5-8. 
All clouds are large enough to be gravitationally resolved and to rely on the Poisson solver, whereby the more massive clouds are spread over much more cells than the lighter ones. 
This is advantageous for AMR grids with respect to particle codes, because the particles density is only derived from pressure equilibrium instead of the cloud's self gravity. 

For the onset of SF, the density plays a crucial role according to our self-regulation recipe \citep{KTH95} what makes the dominant SF in blobs B3 and B4 understandable. Moreover, one must keep in mind that less bound clouds as the larger ones are exposed to the Bernoulli effect, while coming to rest they benefit from the external ICM pressure (Bonnor-Ebert mass) and can be excited to form stars.

Although the gas and stellar masses of the DG rpsLM4 are smaller than of IC\,3418, the occurring SFRs for each starburst in the blobs agree well with the derived SFRs from observations, as e.g. for the blobs of IC\,3418 from $10^{-4} - 10^{-3}\,\solarmass$ yr$^{-1}$. Using the relation between the SFR and $\text{H}\alpha$-luminosity, provided by \cite{KSlaw} to derive the $\text{H}\alpha$-luminosity from the SFR, these star forming blobs would emit luminosities of $L_{\text{H}\alpha} = (0.31 - 2.60)\,\times10^{38}$ erg s$^{-1}$ under the assumption of a filled stellar IMF (but see end of \ref{subsec:sfblobs}).

%%%%%%%%%%%%%%%%%%%%%%%%%%%%%%%%
\section{Conclusions}\label{sec:conclusion}

We have performed a series of DG models which we expose to RP of the IGM while they fall into a cluster environment. The aim is at studying the various effects of RP on DGs. Here we only wish to present a model selection of the most interesting effects which need further elaboration by detailed studies but are already more than exciting. 
Our models achieve the following main results:

\begin{itemize}

\item While the analytical formula of RPS allows to remove the gas from DGs completely, in agreement with observed cluster dEs gas resides in the central galaxy regions. This is caused by gas compression through RP and leads to smaller cloud radii and larger densities, both hampering momentum transfer.

\item For low-mass models with RP the SFR declines more steeply than in isolated models, but jumps up rapidly by two orders of magnitude when the RP exceeds a critical limit.  

\item Weak RP is able to buoy interstellar clouds up from the equatorial plane. When they form stars,  there these young star clusters oscillate about the plane because they are still gravitationally bound and gain a larger velocity dispersion. This effect leads to an ellipsoidal stellar shape with a possibly positive radial blue colour gradient. This assumption could be questioned, however, because at low SFRs the IMF is hardly complete. 

\item Under particular conditions SF can also occur in stripped-off clumps. Obviously, strong RP can push out more massive and dense clumps exceeding $M_J$. 
Smaller RP strips off tenuous clouds or filamentary gas which is incapable to form stars. This dichotomy is verified by observations. Whether Bonnor-Ebert conditions further behind the galaxy allow for SF in the gas or how much environmental processes by means of turbulence, cooling radiation from the ICM and thermal conduction heat the gas and prevent SF, must be the objective of ongoing simulations.

\end{itemize}

%%%%%%%%%%%%%%%%%%%%%%%%%%%%%%%%
\section*{Acknowledgements}

The authors gratefully acknowledge insightful discussions with Thorsten Lisker on Virgo Cluster dEs. We also thank Simone Recchi for discussions on DG evolution and Sylvia Ploeckinger for numerical advice with FLASH. We also acknowledge  the positive report by an anonymous referee and supportive comments which helped to improve the clarity of the paper. 
The software used in this work was in part developed by the DOE NNSA-ASC OASCR Flash Center at the University of Chicago. The computational results presented have been achieved (in part) using the Vienna Scientific Cluster (VSC).

%%%%%%%%%%%%%%%%%%%%%%%%%%%%%%%%%%%%%%%%%%%%%%%%%%



\bibliographystyle{mnras}
\bibliography{rps_papers} % if your bibtex file is called example.bib

%%%%%%%%%%%%%%%%%%%%%%%%%%%%%%%%%%%%%%%%%%%%%%%%%%

%%%%%%%%%%%%%%%%% APPENDICES %%%%%%%%%%%%%%%%%%%%%

%\appendix


%%%%%%%%%%%%%%%%%%%%%%%%%%%%%%%%%%%%%%%%%%%%%%%%%%

% Don't change these lines
\bsp	% typesetting comment
\label{lastpage}
\end{document}